\documentclass[aps,10pt,superscriptaddress,onecolumn,twoside,floatfix,pra,nofootinbib,a4paper]{revtex4-2}

\usepackage[margin=3cm]{geometry}
\usepackage{amsmath,amssymb,amsfonts}
\usepackage{placeins}

\usepackage{graphicx} 
\usepackage{subfig}
\usepackage[colorlinks=true, allcolors=blue, urlcolor=blue]{hyperref}  
\usepackage{amsmath,amssymb,amsthm,bm,amsfonts,mathrsfs,bbm}
\usepackage{xcolor}
\usepackage{comment}
\usepackage{mathtools}
\DeclarePairedDelimiter\bra{\langle}{\rvert}
\DeclarePairedDelimiter\ket{\lvert}{\rangle}
\DeclarePairedDelimiterX\braket[2]{\langle}{\rangle}{#1\,\delimsize\vert\,\mathopen{}#2}

\usepackage{caption}

\newcommand{\ketbra}[1]{\ket{#1}\!\bra{#1}}
\newcommand{\bea}{\begin{eqnarray}}
\newcommand{\eea}{\end{eqnarray}}
\newcommand{\bean}{\begin{eqnarray*}}
\newcommand{\eean}{\end{eqnarray*}}

\newcommand{\id}{\mathbf{1}}
\newcommand{\x}{\mathrm{x}}

\newtheorem*{thm*}{Theorem}

\newtheorem*{lem*}{Lemma}

\newcommand{\tr}{\mathrm{tr}}
\newcommand{\ip}[2]{\langle #1|#2 \rangle}
\newcommand{\G}{\mathcal{G}}
\newcommand{\D}{\mathcal{D}}
\newcommand{\bpm}{\begin{pmatrix}}
\newcommand{\epm}{\end{pmatrix}}

\begin{document}

\title{Sharing quantum indistinguishability with multiple parties } 




\author{Lemieux Wang}
\email{ogisosetsuna@kaist.ac.kr}
\affiliation{School of Electrical Engineering, Korea Advanced Institute of Science and Technology (KAIST), 291 Daehak-ro, Yuseong-gu, Daejeon 34141, Republic of Korea }

\author{Hanwool Lee }
\email{hanwool.h.lee@jyu.fi}
\affiliation{Faculty of Information Technology, University of Jyväskylä,  Finland}

\author{Joonwoo Bae}
\email{joonwoo.bae@kaist.ac.kr}
\affiliation{School of Electrical Engineering, Korea Advanced Institute of Science and Technology (KAIST), 291 Daehak-ro, Yuseong-gu, Daejeon 34141, Republic of Korea }

\author{ Kieran Flatt}
\email{kflatt@kaist.ac.kr}
\affiliation{School of Electrical Engineering, Korea Advanced Institute of Science and Technology (KAIST), 291 Daehak-ro, Yuseong-gu, Daejeon 34141, Republic of Korea }

\begin{abstract}
    Quantum indistinguishability of non-orthogonal quantum states is a valuable resource in quantum information applications such as cryptography and randomness generation. In this article, we present a sequential state-discrimination scheme that enables multiple parties to share quantum uncertainty, in terms of the max-relative entropy, generated by a single party. Our scheme is based upon maximum-confidence measurements and takes advantages of weak measurements to allow a number of parties to perform state discrimination on a single quantum system. We review known sequential state discrimination and show how our scheme would work through a number of examples where ensembles may or may not contain symmetries. Our results will have a role to play in understanding the ultimate limits of sequential information extraction and guide the development of quantum resource sharing in sequential settings. 
\end{abstract}

\maketitle

\section{Introduction}

That non-orthogonal quantum states cannot be perfectly distinguished lies at the heart of quantum theory. This inherent uncertainty in measurements limits detection, but at the same time is a resource that allows for quantum information processing beyond classical limits. Protocols for secure randomness generation and communication, for example, take advantage of this property. There is therefore significant interest, both foundational and practical, in developing tools for understanding how indistinguishability can be distributed. 

In this article, we detail a scheme for sharing quantum indistinguishability in a sequential manner: one party prepares an ensemble and a number of parties measure in turn to extract information. In realistic settings, in which parties are spatially separated and communicate only through photon transmission, it would be impractical for all parties to access preparation devices. They would, instead, have access only to individual optical elements. It has previously been shown that this set-up allows for a number of parties to sequentially share nonlocal correlations, generated from the violation of Bell inequalities, among multiple parties \cite{Silva2015, Colbeck2020}. 

The communication primitive which captures quantum indistinguishability is state discrimination \cite{barnett2009quantum, bergou2010discrimination, bae2015quantum}. One party prepares a quantum system in one state drawn from an a priori ensemble and sends this system to a second party. The latter performs a measurement and aims to determine the chosen state.  The optimisation of this measurement can take the form of a number of strategies, depending on the desired application. Here we focus upon maximum-confidence measurements (MCMs) \cite{croke2006maximum, Mosley2006, lee2022maximum}. While a number of no-go theorems, such as that on cloning, may suggest that multiple parties cannot access in turn information from a single system, our protocol avoids these restrictions through the use of weak measurements.

Our scenario consists of a single quantum system which is repeatedly measured by different parties. We assume that the measuring parties may not communicate classically between themselves but that the experimental settings (i.e., the initial ensemble and implemented measurements) are known to all. The aim of the parties is to determine the initial state. In order to do this, they aim to optimise their confidences while implementing measurements that allow future parties to also access that information. It turns out that all parties can achieve an equally high value of maximum confidence when the positive-operator-valued-measure (POVM) elements describing their conclusive detection events are linearly independent. The result contrasts with sequential Bell violations, where it is weak measurements that each party applies to establish the distribution of nonlocal correlations \cite{PhysRevLett.114.250401, Colbeck2020}.


We consider also sequential maximum-confidence state discrimination of a single quantum system generated in one of a set of linearly dependent states, such as trine qubit states, mirrored symmetric states and geometrically uniform states. In such cases parties performing measurements necessarily have decreasing confidence as the number of detection events increases. It is importantly weak measurements that enable sequential state discrimination, in a similar manner to the sequential nonlocality scenario, in which each party also applies weak measurements that do not allow for a maximal Bell violation \cite{PhysRevLett.114.250401, Colbeck2020}. 

In this work, we show the structure of sequential MCMs as channels that minimally disturb quantum states while probabilistically extracting conclusive detection events that give maximum confidence. We present the quantification of randomness appearing in maximum-confidence discrimination in terms of the min- and max-entropy. We also provide a pedagogical overview of MCMs and their derivations with the approach of convex optimisation. Then, we analyse the relation between state evolution and weak measurements in sequential maximum-confidence discrimination. While sequential MCMs keep parties having strictly smaller values of confidence, there is a single convergent state that, in all cases of our consideration, all states in an ensemble converge to, elucidating the role of weak measurements in sequential MCM.  

This article begins in Section \ref{sec:MCM} with a review of maximum-confidence measurements. We then introduce, in Section \ref{sec:SMCM} a protocol for implementing maximum-confidence measurements in a sequential setting. Two regimes are discussed: one in which all measuring parties can attain an equally high confidence, and one in which the confidence is unevenly distributed. We then move on, in Sections \ref{sec:twomixedstates},  \ref{sec:symmetricstates} and \ref{sec:mirrorsymmetricstates} to a number of examples: noisy states, symmetric ensembles and mirror-symmetric ensembles respectively. The effect of measurement on the ensemble geometry and the latter’s relation to the attainable confidence is emphasised throughout. Finally, we summarise our results and propose a number of applications in Section \ref{sec:conc}.

\section{Maximum-confidence measurement} \label{sec:MCM}

A scenario of quantum state discrimination can be understood as the task of communicating classical messages between two distant parties using quantum states as information carriers. Consider a scenario in which Alice, the sender, chooses a message $x \in [N]$, where $[N]$ denotes a set of natural numbers up to $N$, with \textit{a priori} probability $q_x$. That is, she prepares an ensemble $\mathcal{S}=\{q_x, \rho_x \}_{x=1}^N$. She sends a $d$-dimensional quantum state $\rho_x$ to Bob, whose task is to optimally guess $x$ by making a measurement. 

The optimal strategy depends on the figure of merit of the given information processing task. For instance, one may be interested in minimising the average error probability of guessing, a strategy called minimum-error discrimination \cite{helstrom1967detection}. On the other hand, one may want to discriminate states without any error by admitting some probability of inconclusive outcomes, a strategy called unambiguous discrimination \cite{Ivanovic1987USD,Dieks1988USD,Peres1998USD}. The optimal state discrimination strategies have been extensively studied and have had profound impact on quantum information science, see reviews \cite{ barnett2009quantum, bergou2010discrimination, bae2015quantum}.

A finer figure of merit that constitutes the previously mentioned strategies is \textit{confidence}. Confidence of $x$ is defined as a conditional probability $\mathrm{Prob}_{P|M} (x|x)$ where $P$ and $M$ denote preparation and measurement outcome respectively.  Maximum-confidence measurement (MCM) \cite{croke2006maximum, Mosley2006MCMExp, lee2022maximum}  is a measurement that maximises confidence for all $x$, and is found via the following optimisation,
\bea
C_{x} := \max_{M } \frac{q_{x} \mathrm{Prob}_{M|P} (x|x) }{\mathrm{Prob}_{ M} ( x)} =  \max_{M_{x}} \frac{ q_{x} \tr[\rho_{x} M_{x} ]  }{ \tr[\rho M_{x}] } \label{eq:conf1}
\eea
where $\rho=\sum_{x=1}^N q_x \rho_x$ denotes an average state of the ensemble and $M_x$ a POVM element that represents the measurement outcome $x$. Note that as each $C_x$ is optimised independently, the resultant set of elements may not form a valid POVM. For this reason, as in unambiguous discrimination, MCMs in general also include an inconclusive outcome. Likewise, when $C_x=1$ for all $x$, then the strategy corresponds to unambiguous discrimination. A maximum confidence measurement realises minimum error discrimination when the average confidence over the whole ensemble is maximised. 

\subsection{Entropic quantification of maximum confidence}

The indistinguishability among states of a quantum ensemble may be expressed in terms of the max-relative entropy. This can be seen in the following manner. The optimisation in Eq. \eqref{eq:conf1} is cast as a semi-definite programming (SDP) by introducing a  parameter, $Q_x=\frac{\sqrt{\rho} M_x \sqrt{\rho}}{\tr[\rho M_x ]}$. Then,
\bea
\text{The primal problem: }C_x=\max_{Q_x\geq 0, \tr[Q_x]=1} \tr[\sqrt{\rho}^{-1}q_x\rho_x \sqrt{\rho}^{-1} Q_x] \label{eq:primal}\, .
\eea
The dual problem is derived in Ref. \cite{lee2022maximum} as
\bea
\text{The dual problem: }C_x=\min \lambda : \lambda \id -  \sqrt{\rho}^{-1}q_x\rho_x \sqrt{\rho}^{-1} \geq 0\, . \label{eq:dual}
\eea
One can easily show that the strong duality holds, so the solutions of the primal and dual problems are identical. 
The dual problem is  directly related to the max-relative entropy, also known as the max R\'enyi divergence, $D_\infty(\cdot||\cdot)$, which is defined as \cite{datta2009min, muller2013quantum}
\bea
D_{\mathrm{max}} (\rho||\sigma)=\begin{cases}
\log ||\sqrt{\sigma}^{-1} \rho \sqrt{\sigma}^{-1} ||_\infty , \text{ if } \mathrm{supp}(\rho) \subseteq \mathrm{supp}(\sigma)\\
\infty , \text{ if } \mathrm{supp}(\rho) \not \subseteq \mathrm{supp}(\sigma), 
\end{cases}
\eea
where $||\cdot||_\infty$ denotes an operator norm. The maximum confidence is then  represented in terms of max-relative entropy,
\bea
C_x=q_x 2^{D_{\mathrm{max}} (\rho_x || \rho)}=2^{D_{\mathrm{max}} (q_x \rho_x || \rho)}\, .
\eea
The maximum confidence gives an operational interpretation of the max-relative entropy through the task of state discrimination, capturing how well a single state $\rho_x$
can be maximally distinguished from other states in the ensemble. Note that the max-relative entropy of two probability distributions $P$ and $Q$ is $D_{\mathrm{max}} (P||Q)=\log \sup_i \frac{p_i}{q_i}$ where $p_i$ and $q_i$ are the elements of $P$ and $Q$ \cite{van2014renyi}. 

It is worth mentioning that the guessing probability in minimum-error discrimination provides the operational meaning of min-entropy as the \textit{uncertainty of classical information given quantum side information in a single-shot scenario} \cite{konig2009operational}. The guessing probability is the maximum average distinguishability of the ensemble. Namely, the guessing probability is
\bea
P_{\mathrm{guess}}=\max_{M} \sum_x q_x \tr[\rho_x M_x]=2^{-H_{\mathrm{min}}(X)}
\eea
where $X$ is a random variable about $x$. Both the distinguishability of an individual state $\rho_x$ and that of the entire ensemble are directly linked to quantum entropies.

\subsection{Structure of maximum confidence measurements}
The optimality conditions of an optimisation problem, also known as the Karush-Kuhn-Tucker (KKT) conditions, are necessary and sufficient conditions that the optimal parameters must satisfy. For maximum confidence measurements, these were shown in Ref. \cite{lee2022maximum} to be
\bea
&&C_x \rho=q_x\rho_x + r_x \sigma_x \label{eq:optcon}\\
&&r_x\tr[\sigma_x M_x]=0 \nonumber
\eea
where $r_x \geq 0$, and $\sigma_x$ is a quantum state called the complementary state. These conditions are called the Lagrangian stability and the complementary slackness, respectively. To get more intuition on these conditions, let us divide the first condition by $C_x$, 
\bea
\rho=\mu_x \rho_x+ (1-\mu_x)\sigma_x 
\eea
where $0 \leq \mu_x=\frac{q_x}{C_x} \leq 1$. The interpretation of the optimality conditions is clear; once we find a non-full rank state $\sigma_x$ that forms $\rho$ by convex mixture with $\rho_x$, the optimal POVM element $M_x$ is any operator whose support lies in the kernel of $\sigma_x$. Therefore, the problem of maximum confidence measurement in Eq. (\ref{eq:conf1}) comes down to finding the complementary state $\sigma_x$.  

Denote the spectral decomposition $\sqrt{\rho}^{-1}q_x\rho_x\sqrt{\rho}^{-1}=\sum_{i=1}^d \lambda_x^i \ketbra{\lambda_x^i}$, where $\lambda_x^i$ are the eigenvalues in decreasing order, $C_x=\lambda_x^1\geq \lambda_x^2\geq \ldots \geq \lambda_x^d$. Let $g_x$ denote the degeneracy of the largest eigenvalue, $\lambda_x^1=\lambda_x^2=\ldots=\lambda_x^{g_x}$. The Lagrangian stability condition is
\bea
r_x\sigma_x=\sqrt{\rho} (C_x \id -\sqrt{\rho}^{-1}q_x \rho_x\sqrt{\rho}^{-1} )\sqrt{\rho}
=\sqrt{\rho} \sum_{i>g_x} (C_x-\lambda_x^i)\ketbra{\lambda_x^i})\sqrt{\rho}  \, .
\eea
A set of linearly independent states $\{ \ket{\phi_x^i} \}_{i=1}^{g_x}$, where $\ket{\phi^i_x}=\frac{\sqrt{\rho}^{-1} \ket{\lambda_x^i}}{||\sqrt{\rho}^{-1} \ket{\lambda_x^i}||}$, forms a basis of the kernel of $\sigma_x$.  Therefore, a POVM element of MCM is represented as
\bea\label{eq:povm}
M_x=\sum_{i,j=1}^{g_x} a_x^{ij} \ket{\phi^i_x}\bra{\phi^j_x} \label{eq:mcmgeneral}
\eea
where $a_x^{ij}\geq0$ are constants freely chosen up to the constraint that $\{M_x\}$ forms a valid POVM. 

For any ensemble of quantum states, the largest eigenvalue of the operator $\sqrt{\rho}^{-1}q_x\rho_x\sqrt{\rho}^{-1}$ has a degeneracy of at least one. Therefore, one can always find rank-one POVM elements of MCM,
\bea
M_x=a_x\Pi_x, \label{eq:mcmrrankone}
\eea
where $0 \leq  a_x \leq 1$ and $\Pi_x=\ketbra{\phi_x}$ is a rank-one eigenprojector associated with the largest eigenvalue, which we call the MCM projector. When the state is pure, $\rho_x=\ketbra{\psi_x}$, the MCM projector takes a simple form, $\ket{\phi_x}=\frac{\sqrt{\rho}^{-1}\ket{\psi_x}}{||\sqrt{\rho}^{-1}\ket{\psi_x}||}$.
Note that when $\rho_x$ are qubits then MCM must be rank-one. The structure of qubit MCM has been investigated in Ref. \cite{lee2022maximum}.

The general form of MCM in Eq. (\ref{eq:mcmgeneral}) contains arbitrary constants $a_x^{ij}$, which can be optimised with respect to some figure of merit. Since there might not exist suitable parameters $a_x^{ij}$ such that $\sum_{x=1}^N M_x=\id$, it may be necessary to include an additional outcome, known as the \textit{inconclusive outcome}, represented by an additional POVM element $M_0=\id-\sum_{x=1}^NM_x$. 
The parameters $a_x^{ij}$ are typically optimised to minimise the probability of inconclusive outcomes,
\bea
\eta_0=\tr[\rho M_0] \, .
\eea
The minimum inconclusive rate can be found by an SDP. For instance, when the POVM elements are rank-one as in Eq. (\ref{eq:mcmrrankone}), the optimisation is written as
\bea
\min \eta_0=1-\sum_{x=1}^N  a_x\tr[ \rho  \Pi_x]: a_x \geq 0 ~ \forall x \in [N], \id - \sum_x a_x \Pi_x \geq 0 \, .
\eea
We refer to the MCM that yields the minimum inconclusive rate as the optimal MCM. 

\section{Sequential maximum confidence measurement} \label{sec:SMCM}

\begin{figure}[!t]
    \centering
    \includegraphics[width=\textwidth]{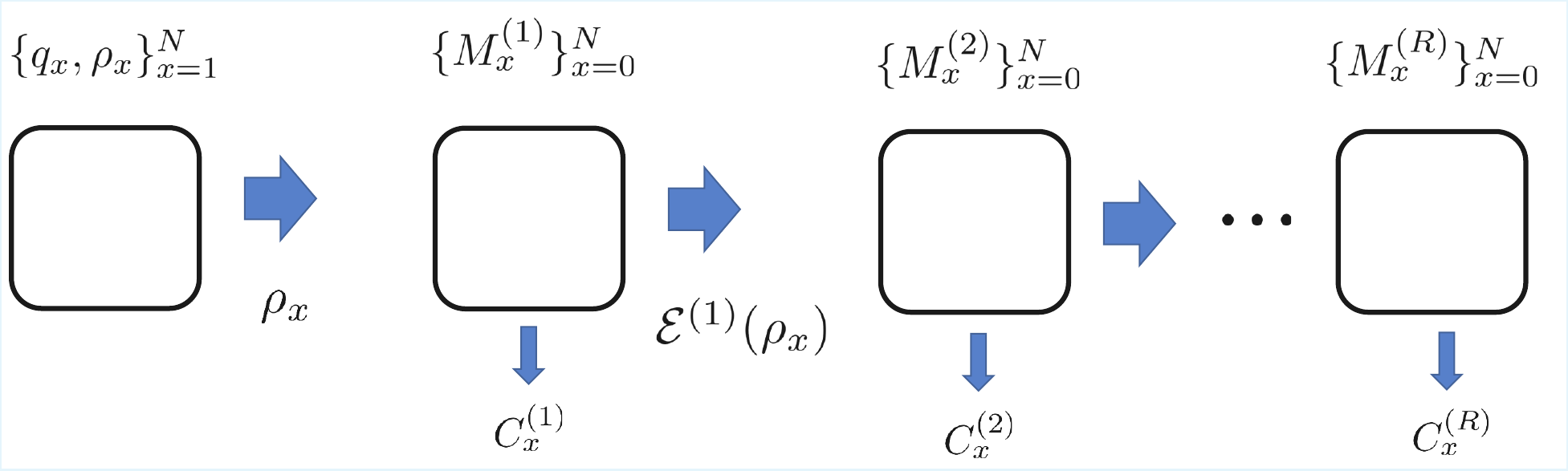}
    \caption{The scenario of sequential maximum confidence measurements. One party prepares a state taken from the ensemble $\{q_x, \rho_x\}_{x=1}^N$. This system is then measured in turn by $R$ parties who each implements an MCM, updating the state to $\mathcal{E}^{(j)}(\rho_x)$ after each measurement, such that their confidence is $C_x^{(j)}$.}
    \label{fig:smcm}
\end{figure}

In this section, we extend the theory of maximum confidence measurement to a multi-party sequential scenario. Suppose there are $R$ parties, denoted as $B_j$ for $j=1,...,R$. Alice prepares an ensemble of states $\{q_x, \rho_x\}_{x=1}^N$ and sends it to the first party, $B_1$. In sequential discrimination, each party aims to guess $x$ by sequentially measuring a single system. Namely, $B_j$ receives a post-measurement state from $B_{j-1}$, measures it, and sends the post measurement state to $B_{j+1}$. 
We assume that classical communication is not allowed but each party has full knowledge of the whole sequential protocol, i.e., the ensemble and the measurements implemented by others. That is, they know which ensemble they receive. The scenario is displayed in Fig. \ref{fig:smcm}.

We begin by considering sequential state discrimination in general. The $j$th receiver's measurement is represented as a quantum channel with Kraus operators 
\bea
\mathcal{E}^{(j)} (\cdot)=\sum_i K^{(j)}_i (\cdot)K_i^{(j)\dag},
\eea
where $\{K_i^{(j)\dag} K_i^{(j)}\}_{i}$ forms a POVM. Since prior probabilities do not change under a channel, the ensemble $\mathcal{S}^{(j)}$ received by $B_j$ is 
\bea
\mathcal{S}^{(j)}=\{q_x, \rho_x^{(j)}\}_{x=1}^N,
\eea
where $\rho_x^{(j)}=\mathcal{E}^{(j-1)} \circ \ldots \circ \mathcal{E}^{(1)} (\rho_x)$ and $\rho_x^{(1)}=\rho_x$. 

\subsection{Figure of merit in sequential state discrimination}

The guessing probability in a single-party scenario can be extended to the joint success probability in a sequential scenario. Let us denote by $P$ the preparation and by $M^{(j)}$ the measurement outcome observed by $B_j$. We define the joint success probability, $P_J$, as the probability that all receivers make the correct guess 
\bea
P_{J}&=&\sum_x q_x \mathrm{Prob}_{M^{(1)},\ldots , M^{(R)}|P} (x,\ldots , x| x) \label{eq:joint}\\
&=&\sum_x q_x \tr[K_x ^{(R-1)} \ldots K_x ^{(1)}\rho^{(1)}_x K_x ^{(1)\dag} \ldots K_x ^{(R-1)\dag} M_x^{(R)}] \nonumber \\
&=&\sum_x q_x \tr[\rho^{(1)}_x \hat{M}^{(R)}_x] \nonumber
\eea
where 
\bea
\hat{M}^{(R)}_x&=&K_x ^{(1)\dag} \ldots K_x ^{(R-1)\dag} M_x^{(R)}K_x ^{(R-1)} \ldots K_x ^{(1)} \, . \label{eq:jointm}
\eea
Likewise, one can define joint probability of inconclusive outcomes,
\bea \label{eq:jointinc}
P_I=\sum_x q_x \mathrm{Prob}_{M^{(1)},\ldots , M^{(R)}|P} (0,\ldots , 0| x)=\sum_x q_x \tr[\rho_x \hat{M}^{(R)}_0]\, .
\eea

Extraction of information from a quantum system necessarily disturbs the state. The information gain $\G^{(j)}$ associated with $B_j$'s measurement is quantified by the guessing probability
\bea
\G^{(j)}=\sum_x q_x \tr[\rho^{(j)}_x M_x^{(j)}].
\eea
Since $\G^{(j)}$ is specified by only the POVM, and not the Kraus operators, one may freely optimise the latter to minimise the measurement disturbance for a fixed information gain. Ideally, we want to choose Kraus operators that give $\G^{(j)}$ and minimally disturb the states so that the maximal amount of information is left in the post-measurement states.

We wish to quantify the disturbance by a distance measure $\D(\mathcal{S}^{(j)},\mathcal{S}^{(j+1)})$ between the pre- and post-measurement ensembles $\mathcal{S}^{(j)}$ and $\mathcal{S}^{(j+1)}$. 
One may consider various distance measures, such as the average fidelity \cite{wilde2013quantum}. The choice of which distance measure to use in practice is dependent upon the particular application, and a full study of different choices is outside the scope of this work. However, a natural choice is the trace distance, defined as
\bea \label{eq:distance}
\D(\mathcal{S}^{(j)},\mathcal{S}^{(j+1)})=\sum_x q_x ||\rho_x^{(j)}-\rho_x^{(j+1)}||_1\, ,
\eea
which arises as a measure of the average success probability in state discrimination \cite{bae2015quantum}.
The optimisation task giving minimally disturbing Kraus operators is then:
\bea
&&\textrm{minimise ~} \D(\mathcal{S}^{(j)},\mathcal{S}^{(j+1)})=\sum_x q_x ||\rho_x^{(j)} - \sum_i K^{(j)}_i \rho^{(j)}_x K^{(j) \dag}_i ||_1\\ 
&&\textrm{subject~to~} \G^{(j)}=\sum_x q_x \tr[\rho^{(j)}_x K_x^{(j)\dag}K_x^{(j)}]  \, . \nonumber
\eea
This optimisation is difficult to solve in general. However, if we assume that the POVM forms a MCM, this problem can be simplifed and solved analytically in certain cases. By using the triangle inequality, a lower bound of $\mathcal{D}$ can be found as
\bea \label{eq:tracelower}
||\rho^{(j)} - \rho^{(j+1)}||_1 \leq \D(\mathcal{S}^{(j)},\mathcal{S}^{(j+1)})
\eea
where $\rho^{(j)}=\sum_x q_x \rho_x^{(j)}. $

\subsection{Sequential MCM with equally high maximum confidence } 
The set of maximum confidences of $B_j$ are written as $C_x^{(j)}$. Since each measurement can be represented as a channel, it holds that
\bea
C_x^{(1)}\geq C_x^{(2)}\geq \ldots \geq C_x^{(j)}, \forall x
\eea
It is natural to ask under what conditions strict inequalities hold. In Ref. \cite{bergou2013extracting}, it is shown that equally high confidence can be achieved in sequential unambiguous discrimination of two pure states, that is, $C_x^{(1)}=\ldots = C_x^{(R)}=1, \forall x$. In Ref. \cite{Lee2025Sequential}, it was shown that equally high confidence can be achieved with linearly independent measurements: 
\\

\textbf{Proposition} \label{prop:amcm} \cite{Lee2025Sequential}. Sequential MCM with equally high confidence can be realised if and only if the POVM elements are linearly independent. 
\\

In many cases, the condition for equally high confidence is equivalent to $d \geq N$, where $d$ is the dimension of the Hilbert space. For instance, sequential unambiguous discrimination of pure states can be implemented if they are linearly independent, as stated in the proposition. On the other hand, if they are linearly dependent, the maximum confidence of the subsequent party must be lower than that of the first party.


\subsection{Sequential MCM with weak measurements}

The above proposition tells us that there exist ensembles for which sequential MCM with equally high confidence is impossible. One may therefore ask how to proceed for the wider range of ensemble for which the MCM is provided by a POVM with linearly dependent elements. The solution is to implement weak measurements. 

If the POVM implementing MCM for a given ensemble is $\{ M_x \}_{x=0}^{N}$ the measurement can be made weak by decreasing the probability of the conclusive outcomes. We have
\bea \label{eq:weakconc}
M_x &\rightarrow \tilde{M}_x = \alpha_x M_x \, \, x \in [N] \\
M_0 &\rightarrow \tilde{M}_0 = \mathbf{1} - \sum_{x=1}^{N} \tilde{M}_x \label{eq:weakinc}
\eea 
where $\{ \alpha_x \}_{x=1}^{N}$ are a set of parameters to be freely chosen. The role of this weakening is to preserve some information in the initial states, allowing for subsequent extraction.

In the simplest case, and the one which we focus on throughout, the set of parameters is chosen to be equal: $\alpha_x = \alpha$ for all $x$. Then, the corresponding change in the inconclusive POVM element takes the simpler form
\bea
\tilde{M}_0 = (1 - \alpha ) \mathbf{1} + \alpha M_0.
\eea

Here, it can be readily seen the parameter $\alpha$ determines the inconclusive outcome rate. Note that changing the conclusive POVM elements in this manner does not change the confidence. Using this POVM the confidences are:
\bea 
\tilde{C}_x = \frac{q_x {\rm Prob}_{M|P}(x|x)}{{\rm Prob}_{M}(x)} = \frac{q_x \tr[\rho_x \alpha M_x ]}{\tr[\rho \alpha M_x]} = \frac{ q_x \tr[\rho_x M_x ]}{\tr[\rho M_x]} = C_x.
\eea
This means that $B_1$ always attains the maximum confidence.

The most general Kraus operators giving the above measurement are
\bea \label{eq:weakkraus}
&&K_x  = \sqrt{\alpha} V_x \sqrt{M_x},  \, \, \forall x \in [N], \\
&&K_0 = V_0 \left( \sqrt{1-\alpha} \mathbf{1} + i \sqrt{\alpha}\sqrt{M_0} \right) \nonumber
\eea 
where $\{ V_x \}_{x=0}^{N}$ is a set of unitary operators which may be freely chosen. In scenarios where $\alpha$ is fixed, the task of sequential state discrimination is therefore to optimise over the set of unitary operators $V_x$ in order to minimise the disturbance to the ensemble, as discussed previously. As the MCM consists of projective POVM elements $\ketbra{\phi_x}$ for the conclusive outcomes, Eq. \eqref{eq:weakkraus} may instead be written as
\bea
K_x = \sqrt{\alpha} \ket{\varphi_x}\bra{\phi_x}
\eea
and the choice is instead in terms of the set of states $\ket{\varphi_x}$.

There is a trade-off between the inconclusive outcome rate of each party, represented by $\alpha$, and the success of later measurements: at $\alpha=1$, the inconclusive outcome rate is minimised but at the cost of maximal disturbance, so that later parties have a lower confidence. At $\alpha=0$, the earlier parties learn nothing but a subsequent party is able to implement optimal MCM. In between these extremes, a range of behaviours are available and the choice will depend upon experimental or task-specific considerations.

\section{Sequential MCM of two mixed states} \label{sec:twomixedstates}
Let us give an example that illustrates sequential MCM in which each party can obtain equally high confidence. Consider an ensemble of two mixed states with apriori probability $q_1=q_2=\frac{1}{2}$,
\bea
\rho_{\x} & = & p \ketbra{\psi_{x}}+\frac{1-p}{2} \id, ~\mathrm{where} \label{eq:mixedtwostates} \\
 \ket{\psi_{x}} &= &\cos\frac{\theta}{2}\ket{0} - (-1)^{x} \sin\frac{\theta}{2}\ket{1}.\nonumber 
 \eea
with $p\in (0,1]$ and $x=1,2$. For this ensemble, unambiguous discrimination cannot be realised. Since the MCM projectors are rank-one, they may be written as $\Pi_x=\ketbra{\phi_x}$ where $\ket{\phi_x}$ is the eigenvector of $\sqrt{\rho}^{-1} \rho_x \sqrt{\rho}^{-1}$ associated with the largest eigenvalue, 
\bea
\ket{\phi_{x}} =   \left( \sqrt{ \frac{1+p\cos\theta }{{2}}}\ket{0} + (-1)^{x-1 }\sqrt{ \frac{1-p\cos\theta}{{2}}}\ket{1} \right). ~~~\label{eq:comp}
\eea
The MCM is described by the following POVM elements,
\bea
&&M_x=a_x \ketbra{\phi_x},~ x=1,2,  \\
&&M_0=\id-M_1-M_2 \nonumber
\eea
where $a_x \geq 0$ are arbitrary constants. This measurement yields the maximum confidence
\bea
C = C_{x} = \frac{1}{2}(1+\frac{p\sin\theta}{\sqrt{1-p^2\cos^2\theta}})\, ,
\eea
which is equal for both states. This measurement process can be represented as a channel $\mathcal{E}(\cdot)=\sum_i K_i (\cdot)K_i^\dag $ with Kraus operators $K_i=V_i \sqrt{M_i}$ where $V_i$ is an arbitrary unitary operator. In general, these unitaries can be arbitrarily chosen if one is only interested in the measurement outcomes. We show here, however, that by carefully choosing unitaries we can realise sequential discrimination protocol in which maximum confidence does not decrease.

\subsection{Sequential discrimination with two parties}
 
Let us first consider sequential discrimination by two parties. Suppose the first party uses Kraus operators of the form
\bea \label{eq:mcmkraus}
K_1 & =& \sqrt{a_1} \ket{\varphi_2^\perp}\bra{{\phi}_{1} },~ K_2 = \sqrt{a_2} \ket{\varphi_1^\perp}\bra{ {\phi}_{2}}, ~\mathrm{and} \nonumber\\
K_0 &=& \sqrt{b_1}\ket{\varphi_2^\perp}\bra{ {\phi}_{1} }+\sqrt{b_2}\ket{\varphi_1^\perp}\bra{ {\phi}_{2} },\label{eq:kraus1}
\eea
for some states $\ket{\phi_x}$ and parameters $a_x, b_x \geq 0$. Note that the states in Eq. \eqref{eq:mixedtwostates} can be written as
\bea
\rho_x=C\ketbra{\phi_{x\oplus 1}^\perp }+(1-C)\ketbra{\phi_{x }^\perp} .
\eea
This structure can be derived from the optimality conditions in Eq. \eqref{eq:optcon} for $x=1,2$. The measurement in Eq. \eqref{eq:mcmkraus} yields the post-measurement state
\bea
\tilde{\rho}_x := \sum_i K_i \rho_x K_i^\dag =C \ketbra{\varphi_{x\oplus 1}^\perp}+(1-C)\ketbra{\varphi_{x}^\perp} \, .
\eea
There exists, therefore, a measurement with POVM elements $\tilde{M}_x=d_x \ketbra{\phi_{x} }$ that outputs a confidence of $C$ for some set of parameters $d_x \geq 0$. One can constructively find measurements with Kraus operators of the form in Eq. \eqref{eq:mcmkraus} for all parties. Therefore, all receivers can obtain confidence $C$.

We have so far shown that for two states an arbitrarily large number of parties can achieve the same confidence $C$. The remaining question is: for what range of parameters $\ket{\varphi_x}, a_x, b_x$ can we realise the Kraus operators in Eq. \eqref{eq:mcmkraus}? It is clear that they cannot be realised for arbitrary parameters. For instance, a channel cannot increase distinguishability, so the Kraus operators must satisfy
$||\rho_1-\rho_2||_1 \geq ||\mathcal{E}(\rho_1-\rho_2)||_1$. 
One constraint on the parameters comes from the requirement that POVM elements form a normalised and complete set:
\bea
\id =\sum_{\x=1,2} K_x^\dag K_x +K_0^\dag K_0 = \sum_{x=1}^2 (a_x+b_x)\ketbra{\phi_x} +\sqrt{b_1 b_2}\langle \varphi_1|\varphi_2\rangle (\ket{\phi_2}\bra{\phi_1}+\ket{\phi_1}\bra{\phi_2})\, .
\eea
Let us first take the inner product on both sides by the same state $\ket{\phi_x^\perp}$, for $x=1,2$. We obtain the first condition
\bea
a_x+b_x=\frac{1}{1-|\ip{\phi_1}{\phi_2}|^2}, ~x=1,2. \label{eq:krausconst1}
\eea
By taking inner product on both sides by different states $\ket{\varphi_1^\perp}$ and $\ket{\varphi_2^\perp}$, and by using the above relation, we obtain
\bea
|\ip{\varphi_1}{\varphi_2}|=f(a_1, a_2)^{-\frac{1}{2}}|\ip{\phi_1}{\phi_2}| \label{eq:krausconst2}
\eea
where
\bea
f(a_1,a_2)=(1-a_1(1-|\ip{\phi_1}{\phi_2}|^2))(1-a_2(1-|\ip{\phi_1}{\phi_2}|^2)) 
\eea
Since $f(a_1,a_2) \leq 1$ and the equality holds if and only if $a_1=a_2=0$, it follows that $|\ip{\phi_1}{\phi_2}|>|\ip{\varphi_1}{\varphi_2}|$. The condition for possible choices of $\ket{\varphi_x}$ is therefore that the states become less distinguishable.

To summarise, one can always implement sequential discrimination with non-decreasing maximum confidence for two mixed states by constructing Kraus operators in Eq. \eqref{eq:mcmkraus}. It remains to be seen how to choose the parameters $a_x$ such that the measurement is minimally disturbing. It should first be noted that, without further constraints, a trivial answer is that if $a_1=a_2=0$ then the post-measurement ensemble is undisturbed. However, in such a case no information is extracted by the first party. 

The set of $a_x$ are in fact directly related to the information gain:
\bea \label{eq:mixinfogain}
\mathcal{G}=\frac{1}{2}C(a_1+a_2)(1-|\langle \phi_1 | \phi_2 \rangle |^2).
\eea
Then, the parameters $a_x$ and $\ket{\phi_x}$ are determined by Eq. \eqref{eq:krausconst1} and Eq. \eqref{eq:krausconst2}. The maximum information gain is attained when $a_1=a_2=\frac{1}{1+|\ip{\phi_1}{\phi_2}|}$ \cite{herzog2012optimized}, yielding 
\bea
\max \mathcal{G}=C(1-|\ip{\phi_1}{\phi_2}|)\, .
\eea
Let us find the least disturbing MCM for the mixed states in Eq. (\ref{eq:mixedtwostates}) that minimises the average trace distance in Eq. \eqref{eq:distance}
\bea
\D=1/2(||\rho_1-\tilde{\rho}_1||_1+||\rho_2-\tilde{\rho}_2||_1) \, ,
\eea
under the constraint that $\mathcal{G}$ in Eq. \eqref{eq:mixinfogain} is fixed. Without loss of generality, we can place the states $\ket{\varphi_1^\perp}$ and  $\ket{\varphi_2^\perp}$ on the $X-Z$ plane symmetric to $Z$-axis. 
We address this optimisation problem using the method of Lagrange multipliers, formulated as follows:
\bea
\mathrm{minimise} &~& |\ip{ \varphi_1}{ \varphi_2}| =f(a_1,a_2)^{-\frac{1}{2}}|\ip{\phi_1}{\phi_2}| \\ 
\mathrm{subject~to} &~& \mathcal{G}=\frac{1}{2}C(a_1+a_2)(1-| \langle \phi_1 | \phi_2 \rangle |^2  )\, . \nonumber
\eea
The Lagrangian is
\bea
\mathcal{L}& = &f(a_1,a_2)^{-\frac{1}{2}} |\ip{\phi_1}{\phi_2}|+ 
 \lambda(\mathcal{G}-\frac{1}{2}C(a_1+a_2)(1-| \langle \phi_1 | \phi_2 \rangle |^2  )) 
\eea
where $\lambda$ is the Lagrangian multiplier. By solving $\frac{\partial \mathcal{L}}{\partial a_x}=0$ for $x=1,2$,  we find that the optimal parameters satisfy $a_1=a_2$. 
Solving $\frac{\partial \mathcal{L}}{\partial \lambda}=0$, we find
\bea
a_1=a_2=\frac{\mathcal{G}}{C(1-|\ip{\phi_1}{\phi_2}|^2)}\, . 
\eea
With this choice of the parameters, the optimal overlap is obtained as,
\bea
|\ip{ \varphi_1}{ \varphi_2}| =(1-\frac{\mathcal{G}}{C} )^{-1}   |\langle \phi_1 | \phi_2 \rangle|.  \label{eq:overlaps}
\eea
The measurement is the least-disturbing MCM when the probability of getting outcomes $x=1$ and $x=2$ are identical. 

\subsection{Sequential discrimination with arbitrary number of parties}

\begin{figure}[!t]
    \centering
\includegraphics[width=0.6\textwidth, trim={6cm 6cm 6cm 6cm}]{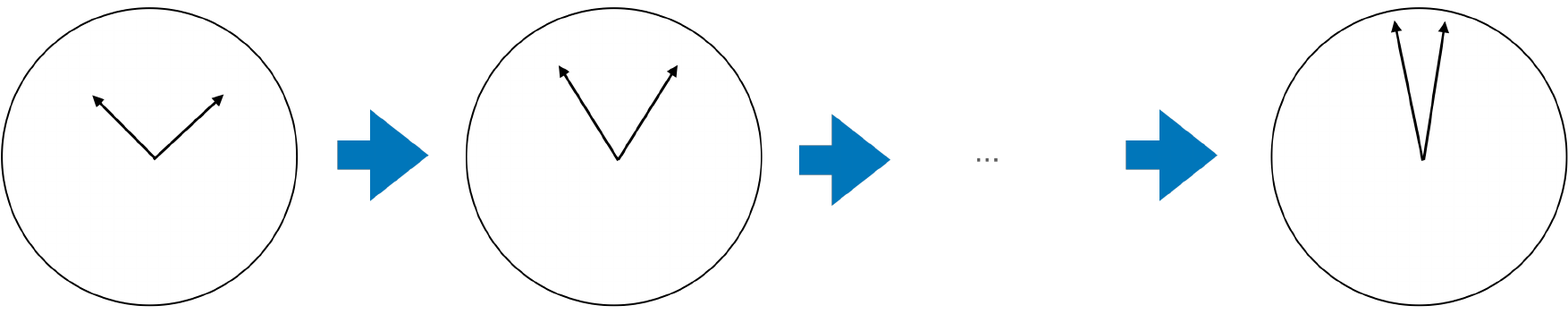}
    \caption{Sequential maximum confidence measurements are applied to an ensemble of two mixed states. The effect is to increase the purity of the states while reducing the angle between them, see Eq. \eqref{eq:mixedsmcmkraus}.}
    \label{fig:twostates}
\end{figure}

We now extend the two-party scenario to a sequential multi-party scenario, 
in which all the parties use the least disturbing Kraus operators. In the sequential scenario, the $j$-th party, $B_j$, receives an ensemble of states
\bea \label{eq:mixedsmcmkraus}
\rho_x^{(j)} =C\ketbra{\phi_{x \oplus 1} ^{(j) \perp}}+(1-C)\ketbra{\phi_{x } ^{(j) \perp}}, x=1,2\, .
\eea
It should be noted that as the states after $B_{j-1}$ measures are also two mixed states, the effect of $B_j$'s measurement is identical: to increase the overlap of the states. Because the maximum confidence remains the same, it can be seen that states' purity increases. The resulting change in the ensemble over multiple measurements is shown in Fig. \ref{fig:twostates}.  The overlap of the states is determined by the information gain of the previous parties using Eq. \eqref{eq:overlaps}, 
\bea
s^{(j)}:= |\ip{\phi_1^{(j)}}{\phi_2^{(j)}}|=\left (1-\frac{\mathcal{G}^{(j-1)}}{C} \right )^{-1}|\ip{\phi_1^{(j-1)}}{\phi_2^{(j-1)}}|=\prod_{k=1}^{j-1} \left (1-\frac{\mathcal{G}^{(k)}}{C} \right )^{-1}|\ip{\phi_1}{\phi_2}|
\eea
where $\mathcal{G}^{(k)}=a^{(k)} C (1-|\ip{\phi_1^{(k)}}{\phi_2^{(k)}}|^2) $ is the information gain by $B_k$ and we take $\ket{\varphi_x^{(1)}}=\ket{\varphi_x}$. Note that the information gain by $B_j$ is upper bounded as $\mathcal{G}^{(j)} \leq C(1-|\ip{\phi_1^{(j)}}{\phi_2^{(j)}}|)$, and when the maximal information is extracted, $s^{(j+1)}=1$, so that subsequent parties cannot attain any information.  
The least disturbing Kraus operators are 
\bea
K_x^{(j)}&=&\sqrt{a^{(j)}} \ket{\phi_{x \oplus 1} ^{(j+1)\perp}}\bra{\phi_x^{(j)}}, x=1,2\\
K_0^{(j)}&=&\sqrt{b^{(j)}} (\ket{\phi_{2} ^{(j+1)\perp}}\bra{\phi_1^{(j)}}+\ket{\phi_{1} ^{(j+1)\perp}}\bra{\phi_2^{(j)}}) \nonumber
\eea
where $b^{(j)}=\frac{1}{1-|\ip{\phi_1^{(j)}}{\phi_2^{(j)}}|}-a^{(j)}$.

Let us find the the joint success probability defined in Eq. \eqref{eq:joint} when $R$ number of parties are involved in the sequential discrimination. This has been investigated in sequential unambiguous discrimination \cite{bergou2013extracting}. The operator of interest is Eq. \eqref{eq:jointm}, 
\bea
\hat{M}^{(R)}_x&=&K_x ^{(1)\dag} \ldots K_x ^{(R-1)\dag} M_x^{(R)}K_x ^{(R-1)} \ldots K_x ^{(1)} = \frac{\prod_{j=1}^R \mathcal{G}^{(j)}}{C ^{R} (1-s^2)} \ketbra{\phi_{x \oplus 1}^\perp} 
\eea
The joint success probability is
\bea
P_J=\frac{1}{2}(\tr[\rho_1 \hat{M}_1^{(R)}]+\tr[\rho_2 \hat{M}_2^{(R)}])=\frac{1}{C^{R-1}} \prod_{j=1}^R \mathcal{G}^{(j)}
\eea
That $s^{R+1}\leq 1$ sets a limit on how much information can be extracted sequentially. To be specific, the information gain $\mathcal{G}^{(j)}$ must satisfy the following inequality,
\bea
s \leq \prod_{j=1}^{R} \left (1-\frac{\mathcal{G}^{(j)}}{C} \right )\, .
\eea
When the final party extracts the maximal information, then equality holds. The optimisation to find the optimal joint success probability is written as follows:
\bea
&&\mathrm{maximise~} P_J=\frac{1}{C^{R-1}} \prod_{j=1}^R \mathcal{G}^{(j)} \\
&&\mathrm{subject~to~} s = \prod_{j=1}^{R} \left (1-\frac{\mathcal{G}^{(j)}}{C} \right ) . \nonumber
\eea
We solve this optimization by the Lagrangian multiplier method. 
The Lagrangian is
\bea
\mathcal{L}(\{\mathcal{G}^{(j)}\},\lambda)=\frac{1}{C^{R-1}} \prod_{j=1}^R \mathcal{G}^{(j)}-\lambda s+\lambda \prod_{j=1}^{R} \left (1-\frac{\mathcal{G}^{(j)}}{C}\right ) 
\eea
where $\lambda$ is a Lagrange multiplier. Since $\nabla \cdot \mathcal{L}=0$,
\bea
\frac{\partial \mathcal{L}}{\partial \mathcal{G}^{(j)}}= \frac{1}{C^{R-1}} \prod_{i \neq j} \mathcal{G}^{(i)}-\frac{\lambda}{C}\prod_{i \neq j} \left (1-\frac{\mathcal{G}^{(j)}}{C}\right )=0, ~ \forall j\, .
\eea
It follows that the information gain must be identical for all parties. Solving the equality constraint, we obtain $\mathcal{G}^{(j)}=C(1-s^{\frac{1}{R}})$, and therefore the optimal joint success probability is
\bea
\max P_J=C(1-s^{\frac{1}{R}})^R
\eea
and the overlaps are
\bea
s^{(j)}=s^{1-\frac{j-1}{R}}, j=1,2,...,R.
\eea

In a similar vein, the joint inconclusive rate can be calculated from Eq. \eqref{eq:jointinc},
\bea
P_I=\tr[\frac{1}{2}(\rho_1+\rho_2)\hat{M}^{(R)}_0]=s.
\eea
This is the same as the optimal inconclusive rate in a single-party scenario \cite{herzog2012optimized}. It tells us that the maximum confidence can be distributed among any number of parties, but also that inconclusive rate must be shared with them.


\section{Sequential MCM of symmetric states} \label{sec:symmetricstates}

In this section, we present a number of examples of sequential MCMs implemented on symmetric ensembles. These use the scheme based on weak measurements. Our emphasis in these examples is on understanding how the geometry of the ensemble is transformed by the measurement.

In each example, our Kraus operators are chosen such that the ensemble's MCM is implemented while the trace distance between pre- and post-measurement ensemble is minimised (see Eq. \eqref{eq:distance}).

\subsection{ Geometrically uniform qubit states} \label{subsec:GUstates}
We begin by studying the ensemble of $N$ pure states symmetrically distributed around a great circle of the Bloch sphere, also known as the geometrically uniform states, so that 
\bea 
\ket{\psi_x} = \frac{1}{\sqrt{2}} \left( \ket{0} + e^{i 2\pi x / N} \ket{1}\right), \, \, \, \forall x \in [N]
\eea
and each state is prepared with equal probability $q_x=1/N$. We note that the average density matrix produced is $\mathbf{1}/2$. Furthermore, the MCM of this ensemble is 
\bea
M_x = \frac{2}{N} \ket{\psi_x}\bra{\psi_x}, \, \, \, \forall x \in [N]
\eea
so that there is no inconclusive outcome, $M_0 = \emptyset$. The maximum confidence for each state in the ensemble is then $C_x = 2/N$.

Let us first characterise this protocol for scenarios with two measuring parties. The MCM does not form a linearly independent set, so we must use weak measurements to make sequential state discrimination possible. The POVM (see Eqs. \eqref{eq:weakconc} and \eqref{eq:weakinc}) is

\bea 
\tilde{M}_x &=& \frac{2\alpha}{N} \ket{\psi_x} \bra{\psi_x}, \, \, \, \forall x \in [N], \\
\tilde{M}_0 &=& (1-\alpha) \mathbf{1}, \nonumber
\eea
with corresponding Kraus operators given by

\bea 
K_x &=& \sqrt{\frac{2 \alpha}{N}} \ket{\varphi_x} \bra{\psi_x}, \,\,\, x \in [N] \\
K_0 &=& \sqrt{1-\alpha} \mathbf{1}, \nonumber
\eea 
and our task is to find the set of vectors $\{ \ket{\varphi_x} \}$ which minimise the distance measure given above in Eq. \eqref{eq:distance}. A short calculation reveals that for this example the distance simplifies to 

\bea
\mathcal{D} = 1 - \sum_{k=1}^N q_x \bra{\psi_x} \tilde{\rho}_x \ket{\psi_x},
\eea
where $\tilde{\rho}_x$ is the average post-measurement state. Minimising this distance is therefore equivalent to maximising the sum on the right hand side. Using the above Kraus operators gives

\bea 
\begin{split}
\sum_{x=1}^N q_x &\bra{\psi_x} \tilde{\rho}_x \ket{\psi_x} = \alpha \left (1- \frac{N}{2} - \frac{1}{N} \sum_{i=1}^N \bra{\varphi_i} \left( \sum_{x=1} q_x \left(1 + \cos{\frac{2\pi(i-x)}{N}} \right) \ket{\psi_x}\bra{\psi_x} \right) \ket{\varphi_i} \right).
\end{split}
\eea
After further simplification and algebraic manipulation, this expression becomes

\bea 
\sum_{x=1}^N q_x &\bra{\psi_x} \tilde{\rho}_x \ket{\psi_x} =\alpha \left( 1 - \frac{N}{2}
- \frac{1}{4} \sum_{i=1}^N \bra{\varphi_i} \left( \ket{\psi_i} \bra{\psi_i} - \ket{\psi_i^{\perp}} \bra{\psi_i^{\perp}} \right) \ket{\varphi_i} \right)
\eea
where $\langle \psi_x |\psi^{\perp}_x \rangle=0$ and, as our aim is to minimise this sum, it can therefore be see seen that the minimally disturbing Kraus operators are rank-one measurements with $\ket{\varphi_k} = \ket{\psi_k}$. 

Let us now use this result to examine the confidence available to each party. The confidence attained by $B_1$ will be $C^{(1)}_x=2/N$, and to find that for $B_2$ we must calculate the post-measurement states. We first note that the coefficient $\alpha$ may be directly related to the inconclusive outcome rate, $\eta_0$, by $
\alpha = 1 - \eta_0$. Using this and the previous result, we can rewrite the Kraus operators above as
\bea 
K_x &=& \sqrt{\frac{2 ( 1 - \eta_0)}{N}} \ket{\psi_x} \bra{\psi_x}, \,\,\, \forall x \in [N] \\
K_0 &=& \sqrt{\eta_0} \mathbf{1}. \nonumber
\eea 
The post-measurement states are 
\bea
\ket{\psi_x} \bra{\psi_x} \rightarrow p_+ \ket{\psi_x} \bra{\psi_x} + p_- \ket{\psi_x^{\perp}} \bra{\psi_x^{\perp}},
\eea
in which 
\bea
p_{\pm} = \frac{1}{2} \left( 1 \pm \frac{1}{2} \left( 1 + \eta_0  \right) \right).
\eea
The combined effect of these operators is therefore to reduce the purity of the states in the ensemble while preserving the angle between the states' Bloch vectors.

\begin{figure}[!t]
    \centering
    \includegraphics[width=0.6\textwidth, trim={6cm 6cm 6cm 6cm}]{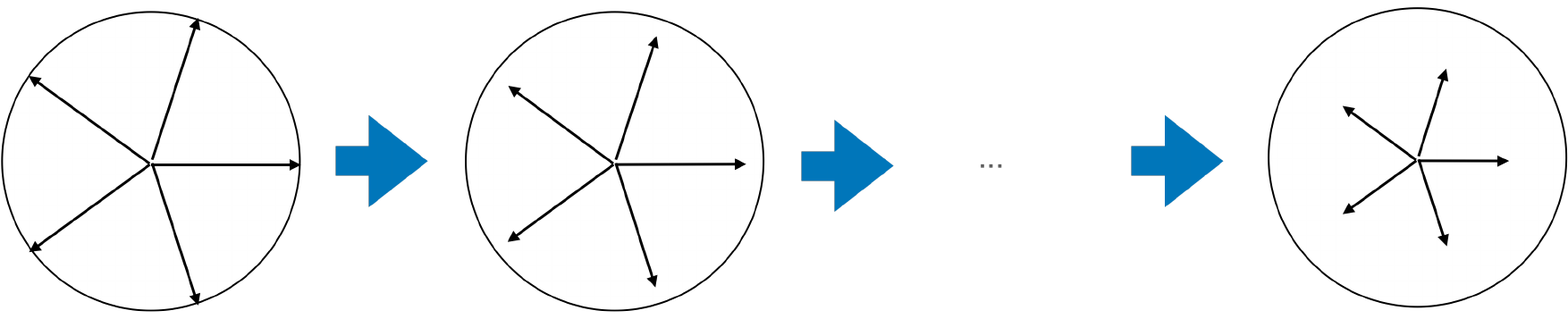}
    \caption{Sequential MCM is implemented on an ensemble of symmetric states. After each measurement, the fixed angle between the states is preserved while the purity is decreased. }
    \label{fig:gustates}
\end{figure}

From the above, we are able to calculate the confidence attained by the second party who measures:
\bea
C_x = \frac{2}{N} p_+.
\eea
It is seen that the subsequent decrease in confidence is given by a factor $p_+$. Interestingly, this factor does not depend on the number of states in the ensemble.

We can extend this analysis to a scenario in which $R$ parties take part in the protocol. It can be seen that the post-measurement ensemble defined by the optimal transformation yields the same maximum confidence measurement as the initial ensemble. This is because both the eigenbasis of each state as well as the average density matrix of the ensemble are unchanged by the optimal choice of Kraus operators. We therefore assign to $B_j$ the Kraus operators 
\bea 
K_x^{(j)} &=& \sqrt{\frac{2 ( 1 - \eta^{(j)}_0)}{N}} \ket{\psi_x} \bra{\psi_x} ,\,\,\, \forall x\in [N] \\
K_0^{(j)} &=& \sqrt{\eta^{(j)}_0} \mathbf{1}. \nonumber
\eea 

Using this, one can show that the state after the first $j$ measurements is
\bea
\rho^{(j)}_x = p^{(j)}_+ \ketbra{\psi_x} + p^{(j)}_- \ketbra{\psi_x^{\perp}},
\eea
in which
\bea
p^{(j)}_{\pm} = \frac{1}{2} \left( 1 \pm \Pi_{k=1}^{(j-1)} \frac{1}{2} ( 1 + \eta_0^{(k)} ) \right).
\eea 
The changing geometry of the states under successive measurements is shown in Fig. \ref{fig:gustates}. We then have 
\bea \label{eq:symmjconf}
C^{(j)}_x = \frac{2}{N} P^{(j)}_+ ,
\eea
for the confidence of $B_j$. Note that $P_+^{(j)} > 1/2$ for all parties. This means that $2P_+^{(j)}>1$, and examination of Eq. \eqref{eq:symmjconf} tells us that all parties have a higher confidence than guessing according to priors. 

\begin{figure}[!t]
    \centering
    \includegraphics[width=0.6\linewidth]{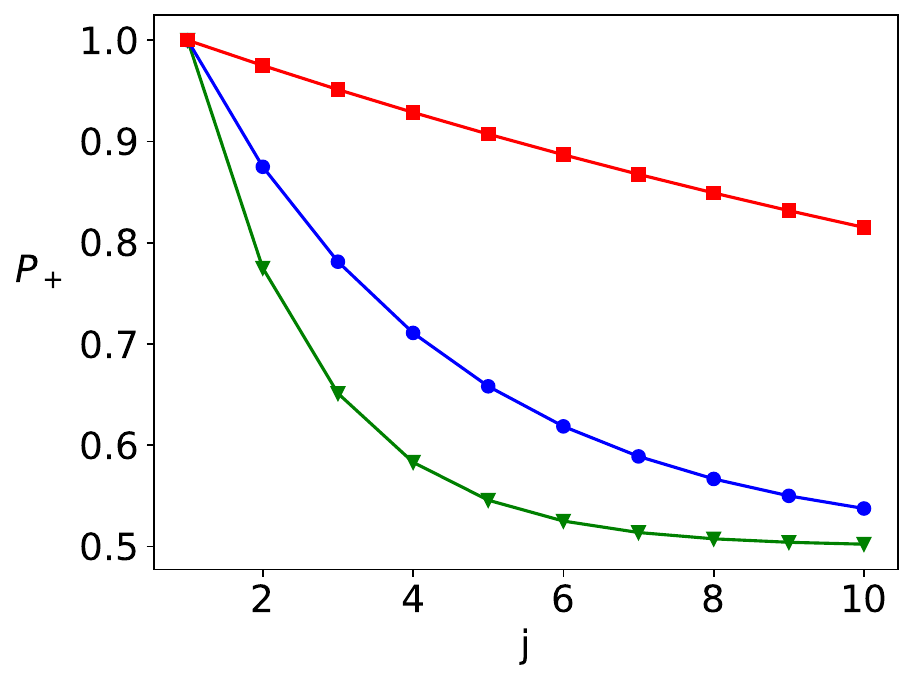}
    \caption{Sequential MCM is implemented on an ensemble of geometrically uniform states. Party $j$ measures a confidence given by $2 P_+ / N$ (see Eq. \eqref{eq:symmjconf}). The factor of proportionality $P_+$ is plotted for three different inconclusive rates: $\eta_0=0.9$ (red squares), $\eta_0=0.5$ (blue dots) and $\eta_0=0.1$ (green triangles).}
    \label{fig:gu_succrate}
\end{figure}

The simplest scenario to look at is the case in which all parties measure with the same inconclusive outcome rate: $\eta^{(j)}_0 = \eta_0$ for all $j$. We have plotted the success rate in Fig. \ref{fig:gu_succrate} for a range of $\eta_0$. It can be seen that the asymptotic limit is that  $\lim_{j\rightarrow \infty } P^{(j)}_+ \rightarrow 1/2$, indicating that as more parties take part in the process their confidence tends to $C_x \rightarrow 1/N$, i.e., guessing according to priors.

\subsection{Lifted geometrically uniform states}

We now consider a class of ensembles of $N$ mixed geometrically uniform states, symmetrically distributed around a lifted plane of the Bloch sphere. These are defined as
\bea \label{eq:mixedgustates}
\rho_x=\lambda \ketbra{\psi_x}+(1-\lambda)\rho =\frac{1}{2} \bpm 
1+\cos\theta & e^{\frac{-2 \pi i x}{N}}\lambda \sin\theta\\
e^{\frac{2 \pi i x}{N}}\lambda \sin \theta & 1-\cos\theta
\epm 
\eea
where $0\leq\lambda\leq1$, $\rho=\frac{1}{N}\sum_x\ketbra{\psi_x}=\frac{1}{N}\sum_x\rho_x$, and 
\bea
\ket{\psi_x}=\cos\frac{\theta}{2}\ket{0}+e^{\frac{2 \pi i x}{N}}\sin\frac{\theta}{2}\ket{1}.
\eea
The reason we consider this class of ensembles is that the post-measurement states always take the form shown in Eq. (\ref{eq:mixedgustates}); that is, the plane on which the states exist is unchanged by sequential MCMs. We begin by studying a scenario with two measuring parties.

The MCM of this ensemble is
\bea
&&M_x=a_x \ketbra{\phi_x}, ~ \forall x\in [N], \\
&&M_0=\id-\sum_x a_x \ketbra{\phi_x}, \nonumber
\eea
where $\ket{\phi_x}=\sin\frac{\theta}{2}\ket{0}+e^{\frac{2 \pi i x}{N}}\cos\frac{\theta}{2}\ket{1}$. The inconclusive rate is lower bounded as $\cos\theta \leq \eta_0$ and the lower bound is achieved by using the parameters $a_x=\frac{2}{N(1+\cos\theta)}, \forall x$ \cite{herzog2012optimized}. Note that the MCM is independent of the value of $\lambda$. The maximum confidence is
\bea \label{eq:liftedgucon}
C_x=\frac{1}{N}(1+\lambda)\, \forall x.
\eea
We begin by calculating the confidence of the second measuring party, $B_2$. The Kraus operators that implement the measurement above can be written as $K_x=V_x \sqrt{M_x}$ where $V_x$ are arbitrary unitary operators.
The Kraus operators for the conclusive parts of MCM for the ensemble in Eq. \eqref{eq:mixedgustates} are
\bea
K_x=\sqrt{a_x}\ket{\varphi_x}\bra{\phi_x},~ \forall x \in [N]\, ,
\eea
where $\ket{\phi_x}=\cos\frac{\phi}{2} \ket{0}+\sin\frac{\phi}{2} \ket{1}$ with $0\leq \phi \leq \frac{\pi}{2}$. Due to symmetry of the states, we take $a=a_x, \forall x$, and it can be shown that $a=\frac{2(1-\eta_0)}{N \sin^2\theta}$.
The Kraus operator for the inconclusive outcome, $K_0$, may also include an arbitrary unitary operator $V_0$,
\bea
K_0=V_0\sqrt{I-\sum_{x=1}^N K_x^\dag K_x}
=V_0\sqrt{I-a\sum_{x=1}^N \ket{\phi_x}\bra{\phi_x}}
=V_0\bpm \sqrt{\frac{\eta_0+\cos\theta}{1+\cos\theta}} & 0\\
0 & \sqrt{\frac{\eta_0-\cos\theta}{1-\cos\theta}}
\epm \, .
\eea
Since $\sqrt{\frac{\eta_0+\cos\theta}{1+\cos\theta}}\geq \sqrt{\frac{\eta_0-\cos\theta}{1-\cos\theta}}$ and the equality holds only when $\eta_0=1$, the operation $\sqrt{M_0}$ transforms the ensemble $\{\rho_x\}$ to be closer to $\ket{0}$. Taking the symmetry of the states into account, we take $V_0=I$ since any other rotation cannot reduce the distance between two ensembles $\{\sqrt{M_0}\rho_x \sqrt{M_0}\}$  and $\{\ket{\psi_x}\}$.
With this choice of Kraus operators, the post-measurement states are
\bea
\tilde{\rho}_x :=\sum_{y=0}^N K_y \rho_x K_y^\dag
=\frac{1}{2}\bpm
1+\cos\tilde{\theta} & e^{-i\frac{2 \pi x}{N}}  \tilde{\lambda} \sin\tilde{\theta}\\
e^{i\frac{2 \pi x}{N}} \tilde{\lambda} \sin\tilde{\theta}& 1-\cos\tilde{\theta}
\epm
\eea
where $\cos\tilde{\theta}=(1-\eta_0)\cos\varphi+\cos\theta$ and 
\bea
\tilde{\lambda}=\lambda \left [ \frac{\frac{1}{2}(1-\eta_0)\sin\varphi+\sqrt{\eta_0^2-\cos^2\theta}}{\sin\tilde{\theta}} \right ]=\lambda \left [ \frac{\frac{1}{2}(1-\eta_0)\sin\varphi+\sqrt{\eta_0^2-\cos^2\theta}}{\sqrt{1-((1-\eta_0)^2\cos^2\varphi + \cos\theta)^2}} \right ].
\eea

\begin{figure}[!t]
    \centering
    \includegraphics[width=0.6\linewidth,  trim={6cm 6cm 6cm 6cm}]{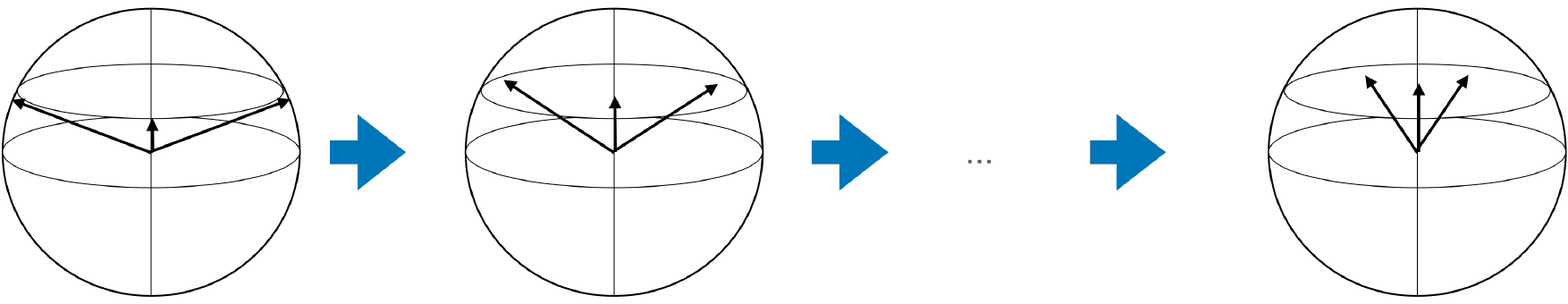}
    \caption{Sequential maximum confidence measurements are applied to three lifted geometrically uniform states, as shown in Eq. \eqref{eq:liftgustates}. The measurement causes the purity of each state to change while their measurement angles with respect to the $Z$ axis are preserved.}
    \label{fig:lifted}
\end{figure}

Let us find the Kraus operators that minimise the average trace distance between $\{\rho_x\}$ and $\{\tilde{\rho}_x\}$,
\bea
\mathcal{D} &=&\frac{1}{N}\sum_x ||\rho_x - \tilde{\rho}_x||_1  \\
&=&\sqrt{(1-\eta_0)^2(1-\sin^2\varphi)+\lambda^2(\frac{1}{2} (1-\eta_0)\sin\varphi +\sqrt{\eta^2_0-\cos^2\theta} -\sin\theta)^2}\, .
\eea
Since $\mathcal{D}$ is monotonically decreasing in terms of $\varphi$, its minimum occurs at $\varphi=\frac{\pi}{2}$. Using the optimal $\varphi$, we obtain $\cos\tilde{\theta}=\cos\theta$ and 
\bea \label{eq:liftedgulamtil}
\tilde{\lambda}=\lambda \Delta~~\mathrm{where}~~\Delta= \left[ \frac{\frac{1}{2}(1-\eta_0)+\sqrt{\eta_0^2-\cos^2\theta}}{\sin\theta} \right ] .\label{eq:deltad}
\eea
Note that $\Delta \leq 1$ and the equality holds if and only if $\theta=\frac{\pi}{2}$
and $\eta_0=1$. The purity of the states, $\tr[\rho_x^2]=\frac{1}{2}(1+\lambda^2\sin^2\theta + \cos^2\theta)$, is decreasing and asymptotically approaches to the purity of $\rho$. 
The post-measurement state can be written as 
\bea \label{eq:liftgustates}
\tilde{\rho}_x=
\tilde{\lambda} \ketbra{\psi_x}+(1-\tilde{\lambda}) \rho
\eea
which shares an identical structure with the initial state in Eq. \eqref{eq:mixedgustates} but with a different parameter $\tilde{\lambda}$. The measurement angle with respect to the $Z$ basis does not change after the measurement, and the states asymptotically approach $\rho$. This change is shown in Fig. \ref{fig:lifted}.
The maximum confidence of the post-measurement state is
\bea
\tilde{C}_x=\frac{1}{N}(1+\tilde{\lambda}),
\eea
where $\tilde{\lambda}$ is expressed in Eq. \eqref{eq:liftedgulamtil}. The resulting confidence above can be compared with Eq. \eqref{eq:liftedgucon}, both of which share a structure.

We can extend these results to construct sequential MCM of the lifted geometrically uniform states by an arbitrary number of parties. We use superscript $j$ to denote the parameters used by $B_j$. Let us denote the inconclusive rate of $B_j$ by  $\eta_0^{(j)}$. Since the structure of the states in Eq. \eqref{eq:mixedgustates} during the whole sequential protocol does not change except the parameter $\lambda$,  
the states that $B_j$ receives are represented as
\bea
\rho^{(j)}_x=\lambda^{(j)} \ketbra{\psi_x}+(1-\lambda^{(j)}) \rho 
\eea
in which we have
\bea
\lambda^{(j)}=\lambda^{(j-1)} \Delta^{(j-1)}=\prod_{k=1}^{j-1} \Delta^{(k)} =\prod_{k=1}^{j-1} \frac{\frac{1}{2}(1-\eta^{(k)}_0)+\sqrt{\eta_0^{(k)2}-\cos^2\theta}}{\sin\theta}
\eea
where we take $\lambda^{(1)}=1$ and $\Delta$ can be found in Eq. (\ref{eq:deltad}). Note that when $\theta=\frac{\pi}{2}$, $\Delta^{(k)}=\frac{1}{2}(1+\eta_0^{(k)})$. 
The minimally disturbing Kraus operators available to $B_j$ are
\bea
K^{(j)}_x=\sqrt{a^{(j)}}\ket{\varphi_x}\bra{\phi_x},
\eea
where $\ket{\varphi_x}=\frac{1}{\sqrt{2}} (\ket{0}+e^{\frac{2 \pi i x}{N}}\ket{1})$, $\ket{\phi_x}= \sin\frac{\theta}{2}\ket{0}+e^{\frac{2 \pi i x}{N}}\cos\frac{\theta}{2}\ket{1}$, and $a^{(j)}=\frac{2(1-\eta^{(j)}_0)}{N \sin^2\theta}$. 
The confidences attained by $B_j$ are given by
\bea
C^{(j)}_x=\frac{1}{N}(1+\lambda^{(j)})=\frac{1}{N}(1+\prod_{k=1}^{j-1} \Delta^{(k)})\, .
\eea
Now consider $R$ receivers, and we want $C_x^{(R)}\geq C_{th}$ for some threshold value of confidence; note that the last party has the smallest value of confidence. When each receiver obtains the same inconclusive rate, $\eta_0^{(j)}=\eta_0, ~\forall j$, the number of parties that can participate in the sequential discrimination can be characterised,
\bea
R\leq 1+\frac{\log(N C_{th} -1)}{\log \Delta}=1+\frac{\log(N C_{th}-1)}{\log(\frac{1}{2}(1-\eta_0)+\sqrt{\eta_0^{2}-\cos^2\theta})-\log \sin\theta} \, .
\eea
One can find that the number of parties relies on detection rates $1-\eta_0$ and the threshold $C_{th}$.

\section{Sequential MCM of mirror-symmetric states} \label{sec:mirrorsymmetricstates}

We now consider sequential maximum confidence measurements applied to three mirror-symmetric states \cite{andersson2002minimum}, given by
\bea
    \begin{split}
        \ket{\psi_1} &= \frac{1}{\sqrt{2}}\left( \ket{0} + \ket{1}\right) \\
        \ket{\psi_2}&=\frac{1}{\sqrt{2}}\left( \ket{0} + e^{i\theta} \ket{1}\right) \\
        \ket{\psi_3} &= \frac{1}{\sqrt{2}} \left( \ket{0} + e^{-i\theta} \ket{1} \right).
    \end{split}
\eea
Note that in the case $\theta=2\pi/3$, this ensemble becomes three geometrically uniform states (see Section \ref{subsec:GUstates}). As sets of states of this kind will reoccur throughout the calculations, we denote by $\mathcal{MS}(\theta)$ any three states with the above structure, where the second and third state are separated by angle $\theta$.  

We begin by studying the case with two measuring parties. It follows from the symmetry of the ensemble that conclusive outcomes of the MCM are given by a set of projectors $\mathcal{MS}(\phi)$. Maximisation of the confidences then gives the relation 
\bea
    \cos(\phi) = \frac{ -4 + \cos(\theta) + 2 \cos(2\theta) + \cos(3\theta)}{6 -2 \cos(\theta) - 4 \cos(2\theta)}.
\eea
between the angle $\theta$ of the states and $\phi$ of the measurement. It is seen that $\theta=2\pi/3$ implies $\phi=2\pi/3$, in agreement with previous results.

We now find the set of $a_j$ for the MCM. Firstly, the symmetry tells us that $a_2=a_3$. From this, the inconclusive outcome rate will be zero if
\bea \label{eq:mirrorweights}
a_1=\frac{-2\cos\phi}{1-\cos\phi}, ~a_2=a_3=\frac{1}{1-\cos\phi},
\eea
which yields maximum confidences
\bea
    C_1 =\frac{1}{2+\cos{\theta}}, ~C_2= C_3 = \frac{3+2\cos{\theta}}{4+ 2\cos{\theta}} .
\eea

Given this MCM, the set of Kraus operators will have the structure $K_x = \sqrt{a_x}\ket{\varphi_x}\bra{\phi_x}$. We now use weak measurements in order to perform MCM for the second party given that the first party has fixed inconclusive outcome rate $\eta_0$. As the minimisation task is in general difficult, we use the lower bound in Eq. \eqref{eq:tracelower} as a distance measure between ensembles. This optimisation gives the set of states $\mathcal{MS}(\varphi)$
in which $\varphi$ is defined by 
\bea
\cos{\varphi}=\frac{1 + 3\cos{\phi}+\cos{\theta}}{3+\cos{\phi}+2\cos{\phi}\cos{\theta}}
\eea
Note that this angle does not depend on $\eta_0$. From this, the maximum confidence of the second party can be calculated. 



We calculate inductively the behaviour of the protocol by solving for both the MCM and minimally disturbing Kraus operators for each party $B_j$. Noting that the effect of measurements is to reduce the purity of each state while reducing the angle between the second and third states, the ensemble after the $j$th measurement can be represented by
\begin{equation}
\begin{split}
    \rho_1^{(j)}&=\frac{1}{2}(\mathbf{1}+r_1^{(j)}\sigma_X)\\
    \rho_2^{(j)}&=\frac{1}{2}(\mathbf{1}+r_2^{(j)}(\cos{(\theta^{(j)})}\sigma_X+\sin{(\theta^{(j)})}\sigma_Y))\\
    \rho_3^{(j)}&=\frac{1}{2}(\mathbf{1}+r_3^{(j)}(\cos{(\theta^{(j)})}\sigma_X-\sin{(\theta^{(j)})}\sigma_Y)),\\
\end{split}
\end{equation}
for some angle $\theta^{(j)}$ and purities $r_x^{(j)}$. Due to the symmetry of the ensemble, $r^{(j)}_2=r^{(j)}_3$ for all $j$, and the initial preparation corresponds to $r_1^{(1)}=r_2^{(1)}=1$. 


Let us begin by finding the MCM for the $j$th party. The conclusive POVM elements have the form
\bea
M^{(j)}_x=a_x^{(j)}\ketbra{\phi_x^{(j)}}
\eea
and are parameterized as a set $\mathcal{MS}(\phi^{(j)})$, where $\phi^{(j)}$ are given by the MCM. These form a complete POVM when $\sum_x a^{(j)}_x \Pi^{(j)}_x=\id$, so that the set $a^{(j)}_x$ are identical to Eq. \eqref{eq:mirrorweights} with $\phi$ replaced by $\phi^{(j)}$.

The corresponding maximum confidence provided by the measurement is 
\bea
    C_1^{(j)}&=&\frac{1+r_1^{(j)}}{3+r_1^{(j)}+2r_2^{(j)}\cos{\theta^{(j)}}}\\
    C_2^{(j)} =C_3^{(j)} &=& \frac{1+r_2^{(j)}\cos{(\phi^{(j)}-\theta^{(j)})}}{3+r_1^{(j)}\cos{\phi^{(j)}}+2r_2^{(j)}\cos{\theta^{(j)}}\cos{\phi}^{(j)}} . \nonumber
\eea
Sequential MCM of the mirror symmetric ensemble can now be performed using a weak measurement (see Eqs. \eqref{eq:weakconc} and \eqref{eq:weakinc}) with fixed inconclusive rate $\eta_0$, whose Kraus operators are defined as
\bea
&&K^{(j)}_x=\sqrt{1-\eta_0^{(j)}} \sqrt{a^{(j)}_x}\ket{\varphi^{(j)}_x}\bra{\phi_x^{(j)}}\label{Kraus_mirror}\\
&&K_0^{(j)}=\sqrt{\eta_0^{(j)}} \id,\nonumber
\eea
where $\ket{\varphi^{(j)}_x}$ are the post-measurement states from the conclusive outcome $x$.

\begin{figure}[!t]
    \centering
\includegraphics[width=0.6\textwidth, trim={5cm 5cm 5cm 5cm}]{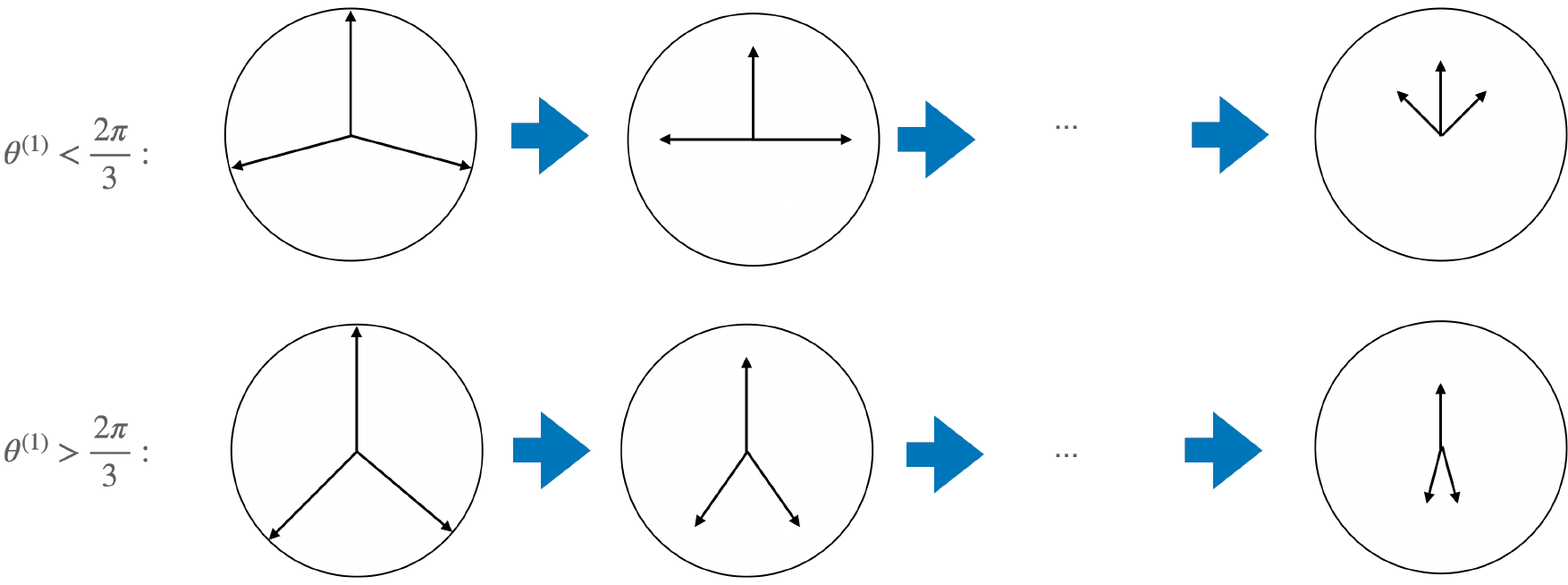}
    \caption{The direction of the mirror-symmetric ensemble evolves depending on the initial angle. Blue arrows represent the MCM channel constructed by optimal Kraus operators Eq. (\ref{Kraus_mirror}).}
    \label{fig:mirror}
\end{figure}

\begin{figure}[!t]
\centering
\subfloat[]{\includegraphics[width=0.4\textwidth]{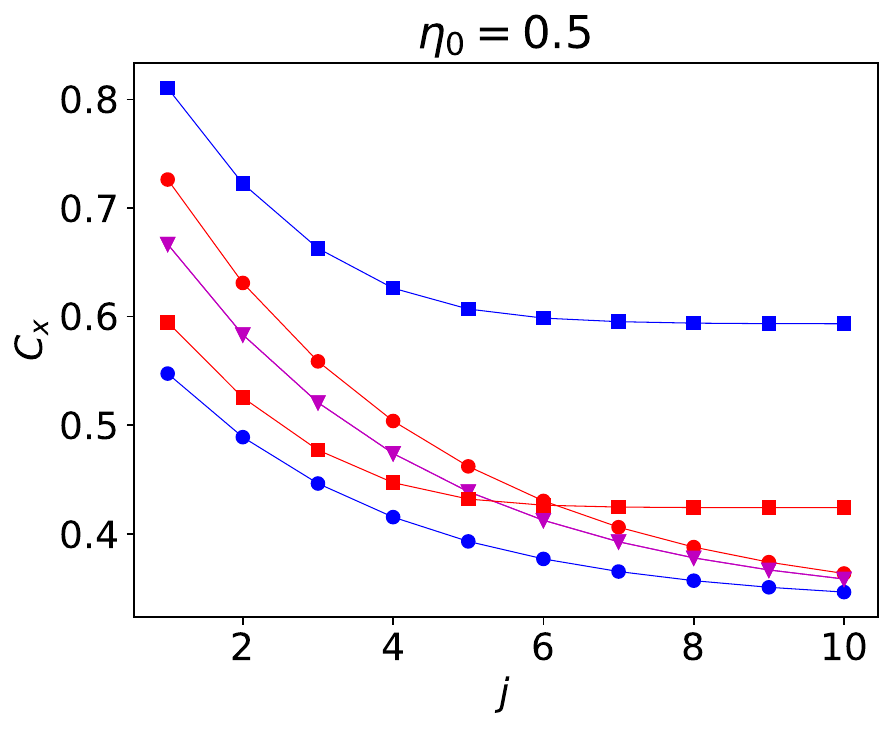}}
\subfloat[]{\includegraphics[width=0.4\textwidth]{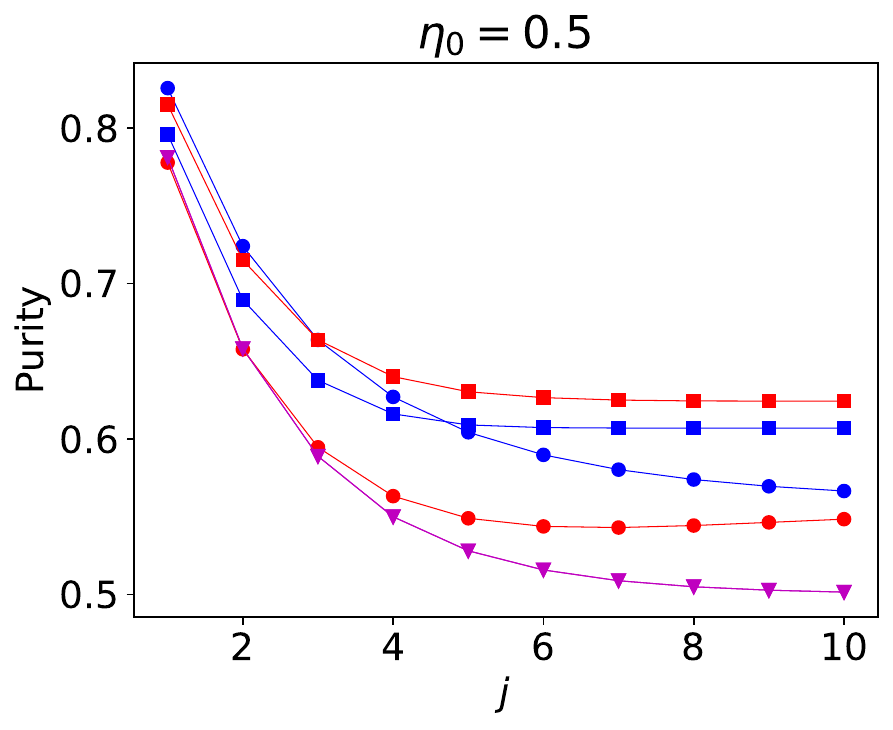}}

\subfloat[]{\includegraphics[width=0.4\textwidth]{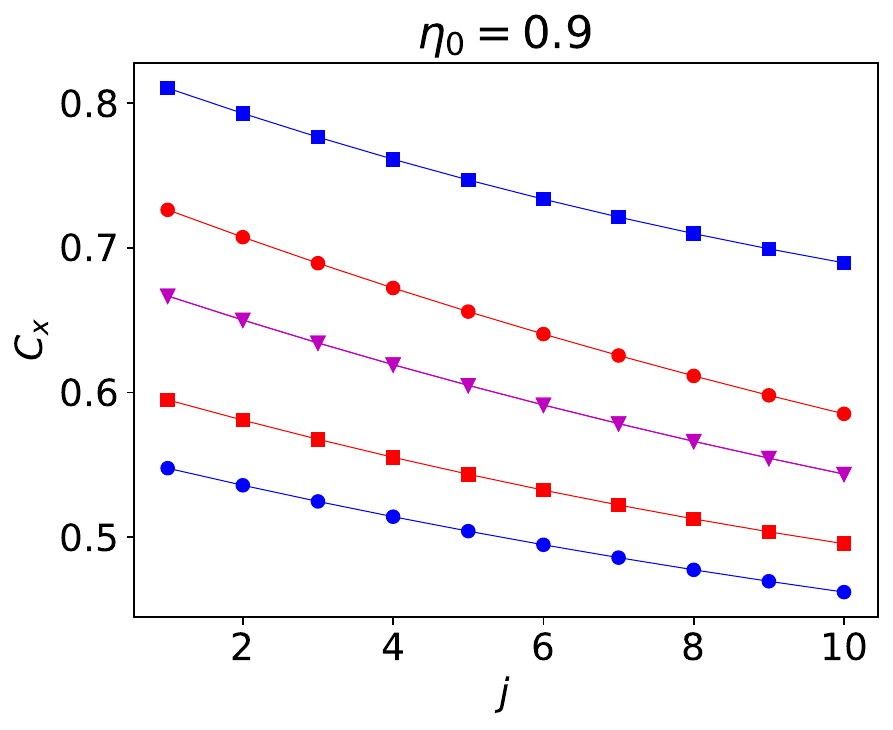}}%
\hspace*{0.005\textwidth}%
\subfloat[]{\includegraphics[width=0.4\textwidth]{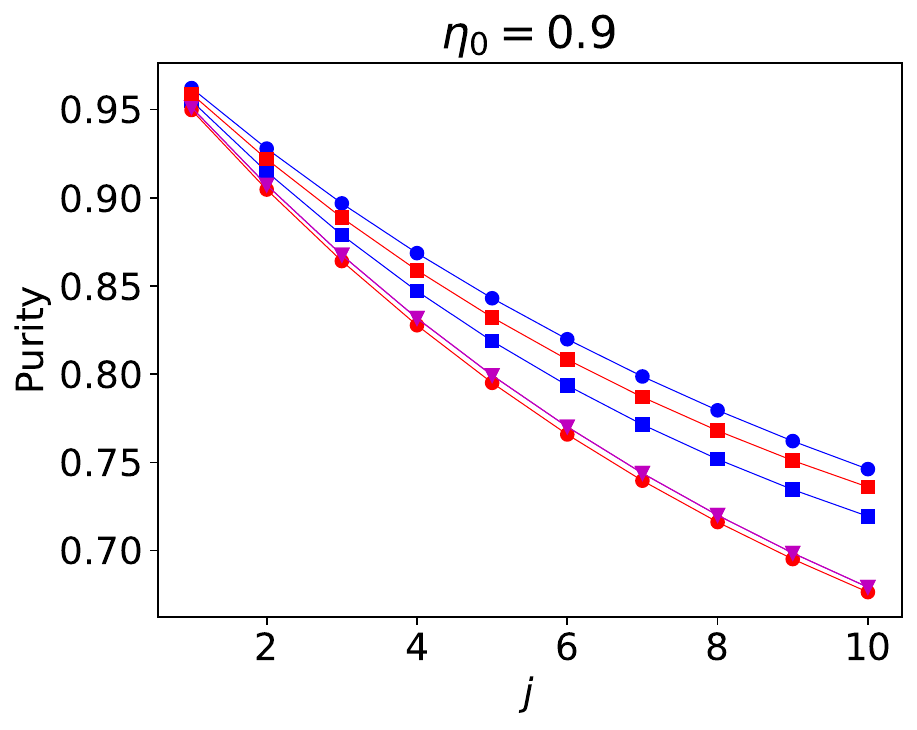}}
\caption{ Numerical results of sequential MCM of the mirror symmetric ensemble. State 1 is coloured blue and states 2 and 3 are red. Initial angles $\theta^{(1)} = 7\pi/9$ (dots), $\theta^{(1)} = 2\pi/3$ (purple triangles) and $\theta^{(1)}=5\pi/9$ (squares) are chosen. The inconclusive rate $\eta_0$ is fixed. For $\eta_0=0.5$, (a) and (b) display the confidences and purity of states respectively. (c) and (d) display these for $\eta_0=0.9$.}
\label{fig:mirror_num}
\end{figure}

We minimise the trace distance with fixed inconclusive outcome rate $\eta_0$ values for all $j$. As this is in general difficult to solve analytically, the minimisation is performed numerically. Fig. \ref{fig:mirror_num} displayed the numerical results for the confidences of party $B_j$ as well as the purity of the ensemble they output. 

We can understand the behaviour of the ensemble under transformation by analytically optimising for the lower bound of trace distance in Eq. \eqref{eq:tracelower}. The angle of the set of states $\mathcal{MS}(\varphi^{(j)})$ characterising the Kraus operators is found as 
\bea
\cos{\varphi^{(j)}}=\frac{3\cos{\phi^{(j)}}+r_1^{(j)}+2r_2^{(j)}\cos{\theta}^{(j)}}{3+r_1^{(j)}\cos{\phi}^{(j)}+2r_2^{(j)}\cos{\theta^{\textcolor{blue}{(j)}}}\cos{\phi}^{(j)}} .
\eea
With this optimised $\varphi^{(j)}$, we can inductively prove the same trend of $\theta^{(j)}$ as shown in the numerical calculation (Fig. \ref{fig:mirror}), depending on initial $\theta^{(1)}$. It can be shown that the following relations hold, for $j\geq 1$,
\bea
\theta^{(j+1)}<\theta^{(j)}~~&\mathrm{for}& ~~\theta^{(1)} < \frac{2\pi}{3},  \nonumber   \\
\theta^{(j+1)}=\theta^{(j)}~~&\mathrm{for}& ~~\theta^{(1)} = \frac{2\pi}{3}, ~~\mathrm{and} \\  
\theta^{(j+1)}>\theta^{(j)} ~~ & \mathrm{for} &~~\theta^{(1)} > \frac{2\pi}{3}. \nonumber 
\eea
Note that the geometrically uniform states are given at $\theta^{(1)}=2\pi/3$, so that the behaviour found here agrees with the case studied above (Sec. \ref{subsec:GUstates}). If the angle is greater than this, the second and third states become closer to the state $(\ket{0}-\ket{1})/\sqrt{2}$ whereas if the angle is closer they approach $(\ket{0}+\ket{1})/\sqrt{2}$.

\section{Conclusions } \label{sec:conc}
To conclude, we have considered sequential maximum-confidence discrimination among multiple parties. Sequential maximum-confidence discrimination can maintain equally high confidence over multiple parties only when an ensemble contains linearly independent states; precisely, rank-one POVM elements for maximum-confidence measurement are linearly independent. 

For ensembles of linearly dependent states, the confidence on conclusive outcomes of parties decreases necessarily. Note also that it is weak measurements that make it possible to realise sequential maximum-confidence measurements over multiple parties. We have considered various examples of linearly dependent states, including ensembles of geometrically uniform states, lifted trine states, and mirror-symmetric states, and investigated the transformation of ensembles over multiple parties. While each party generally makes states less distinguishable in terms of a smaller value of confidence, it turns out that all states converge to a single one, called a convergent state, which may depend on the ensemble given in the beginning. In particular, mirror symmetric states will converge to distinct states depending on an initial condition, i.e., how distinguishable a pair of symmetric states are. Lifted trine states do not converge to an identity state. We have presented a detailed analysis in the relation between state transformation and channels between parties defined by maximum-confidence measurements.

A number of important open questions remain. Here, we have used the trace distance as our measure of disturbance, but it is not clear that this is optimal. One may also consider, for example, the ensemble fidelity instead. It is important to understand how different choices effect the optimality of our scheme. Furthermore, sequential measurements correspond to a process of disturbing states in an ensemble and introduce less confidence on detection events after all. In future investigations, it would be interesting to verify a general convergent state in sequential state discrimination. 

Optical systems are the most natural experimental platform for sequential state discrimination, as the latter is proposed as a communication primitive. Both maximum confidence measurements \cite{Mosley2006MCMExp} and sequential state discrimination \cite{SolisProsser2016} been demonstrated using photons. These implementations may be modified to demonstrate sequential maximum confidence measurements. For realistic scenarios, it is important to understand the robustness of sequential maximum confidence measurements to noise. If both the channel and the states are well-characterised, the maximum confidence measurement may be found. In the simplest case of two pure states, note that the effects of dephasing noise on the measurement can be understood using Eq. \eqref{eq:comp}. It is also seen that the condition for all parties to achieve equal maximum confidence will be unaffected by noise. For more complicated scenarios, a detailed analysis is required.

Finally, we envisage a number of practical applications of our scheme. State discrimination is known to underpin secure randomness extraction \cite{Brask2017, Carceller2022}, and multi-party protocols for generating randomness can be developed based upon our scheme \cite{Carceller2025}. Likewise, our results may be extended to $1$ to $N$ communication settings. In both cases, existing schemes are typically based on nonlocality, which is less experimentally feasible than the prepare-and-measure scenario on which are protocol is based. It is also worth mentioning that state discrimination is known to give an operational characterisation of general resource theories \cite{Takagi2019}. One may therefore also consider extensions of these to the sequential regime.

\section*{acknowledgement}
L.W., J.B. and K.F. are supported by the Institute for Information and Communication Technology Promotion (IITP) (RS-2025-02304540, RS-2025-25464876). H.L. acknowledges financial support from the Business Finland project BEQAH.


\begin{thebibliography}{28}%
\makeatletter
\providecommand \@ifxundefined [1]{%
 \@ifx{#1\undefined}
}%
\providecommand \@ifnum [1]{%
 \ifnum #1\expandafter \@firstoftwo
 \else \expandafter \@secondoftwo
 \fi
}%
\providecommand \@ifx [1]{%
 \ifx #1\expandafter \@firstoftwo
 \else \expandafter \@secondoftwo
 \fi
}%
\providecommand \natexlab [1]{#1}%
\providecommand \enquote  [1]{``#1''}%
\providecommand \bibnamefont  [1]{#1}%
\providecommand \bibfnamefont [1]{#1}%
\providecommand \citenamefont [1]{#1}%
\providecommand \href@noop [0]{\@secondoftwo}%
\providecommand \href [0]{\begingroup \@sanitize@url \@href}%
\providecommand \@href[1]{\@@startlink{#1}\@@href}%
\providecommand \@@href[1]{\endgroup#1\@@endlink}%
\providecommand \@sanitize@url [0]{\catcode `\\12\catcode `\$12\catcode
  `\&12\catcode `\#12\catcode `\^12\catcode `\_12\catcode `\%12\relax}%
\providecommand \@@startlink[1]{}%
\providecommand \@@endlink[0]{}%
\providecommand \url  [0]{\begingroup\@sanitize@url \@url }%
\providecommand \@url [1]{\endgroup\@href {#1}{\urlprefix }}%
\providecommand \urlprefix  [0]{URL }%
\providecommand \Eprint [0]{\href }%
\providecommand \doibase [0]{https://doi.org/}%
\providecommand \selectlanguage [0]{\@gobble}%
\providecommand \bibinfo  [0]{\@secondoftwo}%
\providecommand \bibfield  [0]{\@secondoftwo}%
\providecommand \translation [1]{[#1]}%
\providecommand \BibitemOpen [0]{}%
\providecommand \bibitemStop [0]{}%
\providecommand \bibitemNoStop [0]{.\EOS\space}%
\providecommand \EOS [0]{\spacefactor3000\relax}%
\providecommand \BibitemShut  [1]{\csname bibitem#1\endcsname}%
\let\auto@bib@innerbib\@empty
\bibitem [{\citenamefont {Silva}\ \emph
  {et~al.}(2015{\natexlab{a}})\citenamefont {Silva}, \citenamefont {Gisin},
  \citenamefont {Guryanova},\ and\ \citenamefont {Popescu}}]{Silva2015}%
  \BibitemOpen
  \bibfield  {author} {\bibinfo {author} {\bibfnamefont {R.}~\bibnamefont
  {Silva}}, \bibinfo {author} {\bibfnamefont {N.}~\bibnamefont {Gisin}},
  \bibinfo {author} {\bibfnamefont {Y.}~\bibnamefont {Guryanova}},\ and\
  \bibinfo {author} {\bibfnamefont {S.}~\bibnamefont {Popescu}},\ }\bibfield
  {title} {\bibinfo {title} {Multiple observers can share the nonlocality of
  half of an entangled pair by using optimal weak measurements},\ }\href
  {https://doi.org/10.1103/PhysRevLett.114.250401} {\bibfield  {journal}
  {\bibinfo  {journal} {Phys. Rev. Lett.}\ }\textbf {\bibinfo {volume} {114}},\
  \bibinfo {pages} {250401} (\bibinfo {year} {2015}{\natexlab{a}})}\BibitemShut
  {NoStop}%
\bibitem [{\citenamefont {Brown}\ and\ \citenamefont
  {Colbeck}(2020)}]{Colbeck2020}%
  \BibitemOpen
  \bibfield  {author} {\bibinfo {author} {\bibfnamefont {P.~J.}\ \bibnamefont
  {Brown}}\ and\ \bibinfo {author} {\bibfnamefont {R.}~\bibnamefont
  {Colbeck}},\ }\bibfield  {title} {\bibinfo {title} {Arbitrarily many
  independent observers can share the nonlocality of a single maximally
  entangled qubit pair},\ }\href
  {https://doi.org/10.1103/PhysRevLett.125.090401} {\bibfield  {journal}
  {\bibinfo  {journal} {Phys. Rev. Lett.}\ }\textbf {\bibinfo {volume} {125}},\
  \bibinfo {pages} {090401} (\bibinfo {year} {2020})}\BibitemShut {NoStop}%
\bibitem [{\citenamefont {Barnett}\ and\ \citenamefont
  {Croke}(2009)}]{barnett2009quantum}%
  \BibitemOpen
  \bibfield  {author} {\bibinfo {author} {\bibfnamefont {S.~M.}\ \bibnamefont
  {Barnett}}\ and\ \bibinfo {author} {\bibfnamefont {S.}~\bibnamefont
  {Croke}},\ }\bibfield  {title} {\bibinfo {title} {Quantum state
  discrimination},\ }\href {https://doi.org/10.1364/AOP.1.000238} {\bibfield
  {journal} {\bibinfo  {journal} {Advances in Optics and Photonics}\ }\textbf
  {\bibinfo {volume} {1}},\ \bibinfo {pages} {238} (\bibinfo {year}
  {2009})}\BibitemShut {NoStop}%
\bibitem [{\citenamefont {Bergou}(2010)}]{bergou2010discrimination}%
  \BibitemOpen
  \bibfield  {author} {\bibinfo {author} {\bibfnamefont {J.~A.}\ \bibnamefont
  {Bergou}},\ }\bibfield  {title} {\bibinfo {title} {Discrimination of quantum
  states},\ }\href@noop {} {\bibfield  {journal} {\bibinfo  {journal} {Journal
  of Modern Optics}\ }\textbf {\bibinfo {volume} {57}},\ \bibinfo {pages} {160}
  (\bibinfo {year} {2010})}\BibitemShut {NoStop}%
\bibitem [{\citenamefont {Bae}\ and\ \citenamefont
  {Kwek}(2015)}]{bae2015quantum}%
  \BibitemOpen
  \bibfield  {author} {\bibinfo {author} {\bibfnamefont {J.}~\bibnamefont
  {Bae}}\ and\ \bibinfo {author} {\bibfnamefont {L.-C.}\ \bibnamefont {Kwek}},\
  }\bibfield  {title} {\bibinfo {title} {Quantum state discrimination and its
  applications},\ }\href@noop {} {\bibfield  {journal} {\bibinfo  {journal}
  {Journal of Physics A: Mathematical and Theoretical}\ }\textbf {\bibinfo
  {volume} {48}},\ \bibinfo {pages} {083001} (\bibinfo {year}
  {2015})}\BibitemShut {NoStop}%
\bibitem [{\citenamefont {Croke}\ \emph {et~al.}(2006)\citenamefont {Croke},
  \citenamefont {Andersson}, \citenamefont {Barnett}, \citenamefont {Gilson},\
  and\ \citenamefont {Jeffers}}]{croke2006maximum}%
  \BibitemOpen
  \bibfield  {author} {\bibinfo {author} {\bibfnamefont {S.}~\bibnamefont
  {Croke}}, \bibinfo {author} {\bibfnamefont {E.}~\bibnamefont {Andersson}},
  \bibinfo {author} {\bibfnamefont {S.~M.}\ \bibnamefont {Barnett}}, \bibinfo
  {author} {\bibfnamefont {C.~R.}\ \bibnamefont {Gilson}},\ and\ \bibinfo
  {author} {\bibfnamefont {J.}~\bibnamefont {Jeffers}},\ }\bibfield  {title}
  {\bibinfo {title} {Maximum confidence quantum measurements},\ }\href@noop {}
  {\bibfield  {journal} {\bibinfo  {journal} {Physical review letters}\
  }\textbf {\bibinfo {volume} {96}},\ \bibinfo {pages} {070401} (\bibinfo
  {year} {2006})}\BibitemShut {NoStop}%
\bibitem [{\citenamefont {Mosley}\ \emph
  {et~al.}(2006{\natexlab{a}})\citenamefont {Mosley}, \citenamefont {Croke},
  \citenamefont {Walmsley},\ and\ \citenamefont {Barnett}}]{Mosley2006}%
  \BibitemOpen
  \bibfield  {author} {\bibinfo {author} {\bibfnamefont {P.~J.}\ \bibnamefont
  {Mosley}}, \bibinfo {author} {\bibfnamefont {S.}~\bibnamefont {Croke}},
  \bibinfo {author} {\bibfnamefont {I.~A.}\ \bibnamefont {Walmsley}},\ and\
  \bibinfo {author} {\bibfnamefont {S.~M.}\ \bibnamefont {Barnett}},\
  }\bibfield  {title} {\bibinfo {title} {Experimental realization of maximum
  confidence quantum state discrimination for the extraction of quantum
  information},\ }\href {https://doi.org/10.1103/PhysRevLett.97.193601}
  {\bibfield  {journal} {\bibinfo  {journal} {Phys. Rev. Lett.}\ }\textbf
  {\bibinfo {volume} {97}},\ \bibinfo {pages} {193601} (\bibinfo {year}
  {2006}{\natexlab{a}})}\BibitemShut {NoStop}%
\bibitem [{\citenamefont {Lee}\ \emph {et~al.}(2022)\citenamefont {Lee},
  \citenamefont {Flatt}, \citenamefont {Roch~i Carceller}, \citenamefont
  {Brask},\ and\ \citenamefont {Bae}}]{lee2022maximum}%
  \BibitemOpen
  \bibfield  {author} {\bibinfo {author} {\bibfnamefont {H.}~\bibnamefont
  {Lee}}, \bibinfo {author} {\bibfnamefont {K.}~\bibnamefont {Flatt}}, \bibinfo
  {author} {\bibfnamefont {C.}~\bibnamefont {Roch~i Carceller}}, \bibinfo
  {author} {\bibfnamefont {J.~B.}\ \bibnamefont {Brask}},\ and\ \bibinfo
  {author} {\bibfnamefont {J.}~\bibnamefont {Bae}},\ }\bibfield  {title}
  {\bibinfo {title} {Maximum-confidence measurement for qubit states},\
  }\href@noop {} {\bibfield  {journal} {\bibinfo  {journal} {Physical Review
  A}\ }\textbf {\bibinfo {volume} {106}},\ \bibinfo {pages} {032422} (\bibinfo
  {year} {2022})}\BibitemShut {NoStop}%
\bibitem [{\citenamefont {Silva}\ \emph
  {et~al.}(2015{\natexlab{b}})\citenamefont {Silva}, \citenamefont {Gisin},
  \citenamefont {Guryanova},\ and\ \citenamefont
  {Popescu}}]{PhysRevLett.114.250401}%
  \BibitemOpen
  \bibfield  {author} {\bibinfo {author} {\bibfnamefont {R.}~\bibnamefont
  {Silva}}, \bibinfo {author} {\bibfnamefont {N.}~\bibnamefont {Gisin}},
  \bibinfo {author} {\bibfnamefont {Y.}~\bibnamefont {Guryanova}},\ and\
  \bibinfo {author} {\bibfnamefont {S.}~\bibnamefont {Popescu}},\ }\bibfield
  {title} {\bibinfo {title} {Multiple observers can share the nonlocality of
  half of an entangled pair by using optimal weak measurements},\ }\href
  {https://doi.org/10.1103/PhysRevLett.114.250401} {\bibfield  {journal}
  {\bibinfo  {journal} {Phys. Rev. Lett.}\ }\textbf {\bibinfo {volume} {114}},\
  \bibinfo {pages} {250401} (\bibinfo {year} {2015}{\natexlab{b}})}\BibitemShut
  {NoStop}%
\bibitem [{\citenamefont {Helstrom}(1967)}]{helstrom1967detection}%
  \BibitemOpen
  \bibfield  {author} {\bibinfo {author} {\bibfnamefont {C.~W.}\ \bibnamefont
  {Helstrom}},\ }\bibfield  {title} {\bibinfo {title} {Detection theory and
  quantum mechanics},\ }\href@noop {} {\bibfield  {journal} {\bibinfo
  {journal} {Information and Control}\ }\textbf {\bibinfo {volume} {10}},\
  \bibinfo {pages} {254} (\bibinfo {year} {1967})}\BibitemShut {NoStop}%
\bibitem [{\citenamefont {Ivanovic}(1987)}]{Ivanovic1987USD}%
  \BibitemOpen
  \bibfield  {author} {\bibinfo {author} {\bibfnamefont {I.~D.}\ \bibnamefont
  {Ivanovic}},\ }\bibfield  {title} {\bibinfo {title} {How to differentiate
  between non-orthogonal states},\ }\href
  {https://doi.org/https://doi.org/10.1016/0375-9601(87)90222-2} {\bibfield
  {journal} {\bibinfo  {journal} {Physics Letters A}\ }\textbf {\bibinfo
  {volume} {123}},\ \bibinfo {pages} {257} (\bibinfo {year}
  {1987})}\BibitemShut {NoStop}%
\bibitem [{\citenamefont {Dieks}(1988)}]{Dieks1988USD}%
  \BibitemOpen
  \bibfield  {author} {\bibinfo {author} {\bibfnamefont {D.}~\bibnamefont
  {Dieks}},\ }\bibfield  {title} {\bibinfo {title} {Overlap and
  distinguishability of quantum states},\ }\href
  {https://doi.org/https://doi.org/10.1016/0375-9601(88)90840-7} {\bibfield
  {journal} {\bibinfo  {journal} {Physics Letters A}\ }\textbf {\bibinfo
  {volume} {126}},\ \bibinfo {pages} {303} (\bibinfo {year}
  {1988})}\BibitemShut {NoStop}%
\bibitem [{\citenamefont {Peres}(1988)}]{Peres1998USD}%
  \BibitemOpen
  \bibfield  {author} {\bibinfo {author} {\bibfnamefont {A.}~\bibnamefont
  {Peres}},\ }\bibfield  {title} {\bibinfo {title} {How to differentiate
  between non-orthogonal states},\ }\href
  {https://doi.org/https://doi.org/10.1016/0375-9601(88)91034-1} {\bibfield
  {journal} {\bibinfo  {journal} {Physics Letters A}\ }\textbf {\bibinfo
  {volume} {128}},\ \bibinfo {pages} {19} (\bibinfo {year} {1988})}\BibitemShut
  {NoStop}%
\bibitem [{\citenamefont {Mosley}\ \emph
  {et~al.}(2006{\natexlab{b}})\citenamefont {Mosley}, \citenamefont {Croke},
  \citenamefont {Walmsley},\ and\ \citenamefont {Barnett}}]{Mosley2006MCMExp}%
  \BibitemOpen
  \bibfield  {author} {\bibinfo {author} {\bibfnamefont {P.~J.}\ \bibnamefont
  {Mosley}}, \bibinfo {author} {\bibfnamefont {S.}~\bibnamefont {Croke}},
  \bibinfo {author} {\bibfnamefont {I.~A.}\ \bibnamefont {Walmsley}},\ and\
  \bibinfo {author} {\bibfnamefont {S.~M.}\ \bibnamefont {Barnett}},\
  }\bibfield  {title} {\bibinfo {title} {Experimental realization of maximum
  confidence quantum state discrimination for the extraction of quantum
  information},\ }\href {https://doi.org/10.1103/PhysRevLett.97.193601}
  {\bibfield  {journal} {\bibinfo  {journal} {Phys. Rev. Lett.}\ }\textbf
  {\bibinfo {volume} {97}},\ \bibinfo {pages} {193601} (\bibinfo {year}
  {2006}{\natexlab{b}})}\BibitemShut {NoStop}%
\bibitem [{\citenamefont {Datta}(2009)}]{datta2009min}%
  \BibitemOpen
  \bibfield  {author} {\bibinfo {author} {\bibfnamefont {N.}~\bibnamefont
  {Datta}},\ }\bibfield  {title} {\bibinfo {title} {Min-and max-relative
  entropies and a new entanglement monotone},\ }\href@noop {} {\bibfield
  {journal} {\bibinfo  {journal} {IEEE Transactions on Information Theory}\
  }\textbf {\bibinfo {volume} {55}},\ \bibinfo {pages} {2816} (\bibinfo {year}
  {2009})}\BibitemShut {NoStop}%
\bibitem [{\citenamefont {M{\"u}ller-Lennert}\ \emph
  {et~al.}(2013)\citenamefont {M{\"u}ller-Lennert}, \citenamefont {Dupuis},
  \citenamefont {Szehr}, \citenamefont {Fehr},\ and\ \citenamefont
  {Tomamichel}}]{muller2013quantum}%
  \BibitemOpen
  \bibfield  {author} {\bibinfo {author} {\bibfnamefont {M.}~\bibnamefont
  {M{\"u}ller-Lennert}}, \bibinfo {author} {\bibfnamefont {F.}~\bibnamefont
  {Dupuis}}, \bibinfo {author} {\bibfnamefont {O.}~\bibnamefont {Szehr}},
  \bibinfo {author} {\bibfnamefont {S.}~\bibnamefont {Fehr}},\ and\ \bibinfo
  {author} {\bibfnamefont {M.}~\bibnamefont {Tomamichel}},\ }\bibfield  {title}
  {\bibinfo {title} {On quantum r{\'e}nyi entropies: A new generalization and
  some properties},\ }\href@noop {} {\bibfield  {journal} {\bibinfo  {journal}
  {Journal of Mathematical Physics}\ }\textbf {\bibinfo {volume} {54}}
  (\bibinfo {year} {2013})}\BibitemShut {NoStop}%
\bibitem [{\citenamefont {Van~Erven}\ and\ \citenamefont
  {Harremos}(2014)}]{van2014renyi}%
  \BibitemOpen
  \bibfield  {author} {\bibinfo {author} {\bibfnamefont {T.}~\bibnamefont
  {Van~Erven}}\ and\ \bibinfo {author} {\bibfnamefont {P.}~\bibnamefont
  {Harremos}},\ }\bibfield  {title} {\bibinfo {title} {R{\'e}nyi divergence and
  kullback-leibler divergence},\ }\href@noop {} {\bibfield  {journal} {\bibinfo
   {journal} {IEEE Transactions on Information Theory}\ }\textbf {\bibinfo
  {volume} {60}},\ \bibinfo {pages} {3797} (\bibinfo {year}
  {2014})}\BibitemShut {NoStop}%
\bibitem [{\citenamefont {Konig}\ \emph {et~al.}(2009)\citenamefont {Konig},
  \citenamefont {Renner},\ and\ \citenamefont
  {Schaffner}}]{konig2009operational}%
  \BibitemOpen
  \bibfield  {author} {\bibinfo {author} {\bibfnamefont {R.}~\bibnamefont
  {Konig}}, \bibinfo {author} {\bibfnamefont {R.}~\bibnamefont {Renner}},\ and\
  \bibinfo {author} {\bibfnamefont {C.}~\bibnamefont {Schaffner}},\ }\bibfield
  {title} {\bibinfo {title} {The operational meaning of min-and max-entropy},\
  }\href@noop {} {\bibfield  {journal} {\bibinfo  {journal} {IEEE Transactions
  on Information theory}\ }\textbf {\bibinfo {volume} {55}},\ \bibinfo {pages}
  {4337} (\bibinfo {year} {2009})}\BibitemShut {NoStop}%
\bibitem [{\citenamefont {Wilde}(2013)}]{wilde2013quantum}%
  \BibitemOpen
  \bibfield  {author} {\bibinfo {author} {\bibfnamefont {M.}~\bibnamefont
  {Wilde}},\ }\href {https://books.google.co.kr/books?id=T36v2Sp7DnIC} {\emph
  {\bibinfo {title} {Quantum Information Theory}}},\ Quantum Information
  Theory\ (\bibinfo  {publisher} {Cambridge University Press},\ \bibinfo {year}
  {2013})\BibitemShut {NoStop}%
\bibitem [{\citenamefont {Bergou}\ \emph {et~al.}(2013)\citenamefont {Bergou},
  \citenamefont {Feldman},\ and\ \citenamefont
  {Hillery}}]{bergou2013extracting}%
  \BibitemOpen
  \bibfield  {author} {\bibinfo {author} {\bibfnamefont {J.}~\bibnamefont
  {Bergou}}, \bibinfo {author} {\bibfnamefont {E.}~\bibnamefont {Feldman}},\
  and\ \bibinfo {author} {\bibfnamefont {M.}~\bibnamefont {Hillery}},\
  }\bibfield  {title} {\bibinfo {title} {Extracting information from a qubit by
  multiple observers: Toward a theory of sequential state discrimination},\
  }\href@noop {} {\bibfield  {journal} {\bibinfo  {journal} {Physical review
  letters}\ }\textbf {\bibinfo {volume} {111}},\ \bibinfo {pages} {100501}
  (\bibinfo {year} {2013})}\BibitemShut {NoStop}%
\bibitem [{\citenamefont {Lee}\ \emph {et~al.}(2025)\citenamefont {Lee},
  \citenamefont {Flatt},\ and\ \citenamefont {Bae}}]{Lee2025Sequential}%
  \BibitemOpen
  \bibfield  {author} {\bibinfo {author} {\bibfnamefont {H.}~\bibnamefont
  {Lee}}, \bibinfo {author} {\bibfnamefont {K.}~\bibnamefont {Flatt}},\ and\
  \bibinfo {author} {\bibfnamefont {J.}~\bibnamefont {Bae}},\ }\bibfield
  {title} {\bibinfo {title} {Sequential quantum maximum-confidence
  discrimination},\ }\href {https://doi.org/10.1103/wm73-g73w} {\bibfield
  {journal} {\bibinfo  {journal} {Phys. Rev. A}\ }\textbf {\bibinfo {volume}
  {112}},\ \bibinfo {pages} {052206} (\bibinfo {year} {2025})}\BibitemShut
  {NoStop}%
\bibitem [{\citenamefont {Herzog}(2012)}]{herzog2012optimized}%
  \BibitemOpen
  \bibfield  {author} {\bibinfo {author} {\bibfnamefont {U.}~\bibnamefont
  {Herzog}},\ }\bibfield  {title} {\bibinfo {title} {Optimized
  maximum-confidence discrimination of n mixed quantum states and application
  to symmetric states},\ }\href@noop {} {\bibfield  {journal} {\bibinfo
  {journal} {Physical Review A—Atomic, Molecular, and Optical Physics}\
  }\textbf {\bibinfo {volume} {85}},\ \bibinfo {pages} {032312} (\bibinfo
  {year} {2012})}\BibitemShut {NoStop}%
\bibitem [{\citenamefont {Andersson}\ \emph {et~al.}(2002)\citenamefont
  {Andersson}, \citenamefont {Barnett}, \citenamefont {Gilson},\ and\
  \citenamefont {Hunter}}]{andersson2002minimum}%
  \BibitemOpen
  \bibfield  {author} {\bibinfo {author} {\bibfnamefont {E.}~\bibnamefont
  {Andersson}}, \bibinfo {author} {\bibfnamefont {S.~M.}\ \bibnamefont
  {Barnett}}, \bibinfo {author} {\bibfnamefont {C.~R.}\ \bibnamefont
  {Gilson}},\ and\ \bibinfo {author} {\bibfnamefont {K.}~\bibnamefont
  {Hunter}},\ }\bibfield  {title} {\bibinfo {title} {Minimum-error
  discrimination between three mirror-symmetric states},\ }\href@noop {}
  {\bibfield  {journal} {\bibinfo  {journal} {Physical Review A}\ }\textbf
  {\bibinfo {volume} {65}},\ \bibinfo {pages} {052308} (\bibinfo {year}
  {2002})}\BibitemShut {NoStop}%
\bibitem [{\citenamefont {Sol\'{\i}s-Prosser}\ \emph
  {et~al.}(2016)\citenamefont {Sol\'{\i}s-Prosser}, \citenamefont {Gonz\'alez},
  \citenamefont {Fuenzalida}, \citenamefont {G\'omez}, \citenamefont {Xavier},
  \citenamefont {Delgado},\ and\ \citenamefont {Lima}}]{SolisProsser2016}%
  \BibitemOpen
  \bibfield  {author} {\bibinfo {author} {\bibfnamefont {M.~A.}\ \bibnamefont
  {Sol\'{\i}s-Prosser}}, \bibinfo {author} {\bibfnamefont {P.}~\bibnamefont
  {Gonz\'alez}}, \bibinfo {author} {\bibfnamefont {J.}~\bibnamefont
  {Fuenzalida}}, \bibinfo {author} {\bibfnamefont {S.}~\bibnamefont {G\'omez}},
  \bibinfo {author} {\bibfnamefont {G.~B.}\ \bibnamefont {Xavier}}, \bibinfo
  {author} {\bibfnamefont {A.}~\bibnamefont {Delgado}},\ and\ \bibinfo {author}
  {\bibfnamefont {G.}~\bibnamefont {Lima}},\ }\bibfield  {title} {\bibinfo
  {title} {Experimental multiparty sequential state discrimination},\ }\href
  {https://doi.org/10.1103/PhysRevA.94.042309} {\bibfield  {journal} {\bibinfo
  {journal} {Phys. Rev. A}\ }\textbf {\bibinfo {volume} {94}},\ \bibinfo
  {pages} {042309} (\bibinfo {year} {2016})}\BibitemShut {NoStop}%
\bibitem [{\citenamefont {Brask}\ \emph {et~al.}(2017)\citenamefont {Brask},
  \citenamefont {Martin}, \citenamefont {Esposito}, \citenamefont {Houlmann},
  \citenamefont {Bowles}, \citenamefont {Zbinden},\ and\ \citenamefont
  {Brunner}}]{Brask2017}%
  \BibitemOpen
  \bibfield  {author} {\bibinfo {author} {\bibfnamefont {J.~B.}\ \bibnamefont
  {Brask}}, \bibinfo {author} {\bibfnamefont {A.}~\bibnamefont {Martin}},
  \bibinfo {author} {\bibfnamefont {W.}~\bibnamefont {Esposito}}, \bibinfo
  {author} {\bibfnamefont {R.}~\bibnamefont {Houlmann}}, \bibinfo {author}
  {\bibfnamefont {J.}~\bibnamefont {Bowles}}, \bibinfo {author} {\bibfnamefont
  {H.}~\bibnamefont {Zbinden}},\ and\ \bibinfo {author} {\bibfnamefont
  {N.}~\bibnamefont {Brunner}},\ }\bibfield  {title} {\bibinfo {title}
  {Megahertz-rate semi-device-independent quantum random number generators
  based on unambiguous state discrimination},\ }\href
  {https://doi.org/10.1103/PhysRevApplied.7.054018} {\bibfield  {journal}
  {\bibinfo  {journal} {Phys. Rev. Appl.}\ }\textbf {\bibinfo {volume} {7}},\
  \bibinfo {pages} {054018} (\bibinfo {year} {2017})}\BibitemShut {NoStop}%
\bibitem [{\citenamefont {Roch~i Carceller}\ \emph {et~al.}(2022)\citenamefont
  {Roch~i Carceller}, \citenamefont {Flatt}, \citenamefont {Lee}, \citenamefont
  {Bae},\ and\ \citenamefont {Brask}}]{Carceller2022}%
  \BibitemOpen
  \bibfield  {author} {\bibinfo {author} {\bibfnamefont {C.}~\bibnamefont
  {Roch~i Carceller}}, \bibinfo {author} {\bibfnamefont {K.}~\bibnamefont
  {Flatt}}, \bibinfo {author} {\bibfnamefont {H.}~\bibnamefont {Lee}}, \bibinfo
  {author} {\bibfnamefont {J.}~\bibnamefont {Bae}},\ and\ \bibinfo {author}
  {\bibfnamefont {J.~B.}\ \bibnamefont {Brask}},\ }\bibfield  {title} {\bibinfo
  {title} {Quantum vs noncontextual semi-device-independent randomness
  certification},\ }\href {https://doi.org/10.1103/PhysRevLett.129.050501}
  {\bibfield  {journal} {\bibinfo  {journal} {Phys. Rev. Lett.}\ }\textbf
  {\bibinfo {volume} {129}},\ \bibinfo {pages} {050501} (\bibinfo {year}
  {2022})}\BibitemShut {NoStop}%
\bibitem [{\citenamefont {Carceller}\ \emph {et~al.}(2025)\citenamefont
  {Carceller}, \citenamefont {Lee}, \citenamefont {Brask}, \citenamefont
  {Flatt},\ and\ \citenamefont {Bae}}]{Carceller2025}%
  \BibitemOpen
  \bibfield  {author} {\bibinfo {author} {\bibfnamefont {C.~R.~I.}\
  \bibnamefont {Carceller}}, \bibinfo {author} {\bibfnamefont {H.}~\bibnamefont
  {Lee}}, \bibinfo {author} {\bibfnamefont {J.~B.}\ \bibnamefont {Brask}},
  \bibinfo {author} {\bibfnamefont {K.}~\bibnamefont {Flatt}},\ and\ \bibinfo
  {author} {\bibfnamefont {J.}~\bibnamefont {Bae}},\ }\href
  {https://arxiv.org/abs/2510.19445} {\bibinfo {title} {Sequential
  semi-device-independent quantum randomness certification}} (\bibinfo {year}
  {2025}),\ \Eprint {https://arxiv.org/abs/2510.19445} {arXiv:2510.19445
  [quant-ph]} \BibitemShut {NoStop}%
\bibitem [{\citenamefont {Takagi}\ and\ \citenamefont
  {Regula}(2019)}]{Takagi2019}%
  \BibitemOpen
  \bibfield  {author} {\bibinfo {author} {\bibfnamefont {R.}~\bibnamefont
  {Takagi}}\ and\ \bibinfo {author} {\bibfnamefont {B.}~\bibnamefont
  {Regula}},\ }\bibfield  {title} {\bibinfo {title} {General resource theories
  in quantum mechanics and beyond: Operational characterization via
  discrimination tasks},\ }\href {https://doi.org/10.1103/PhysRevX.9.031053}
  {\bibfield  {journal} {\bibinfo  {journal} {Phys. Rev. X}\ }\textbf {\bibinfo
  {volume} {9}},\ \bibinfo {pages} {031053} (\bibinfo {year}
  {2019})}\BibitemShut {NoStop}%
\end{thebibliography}
\end{document}